\documentclass[aps,prl,twocolumn,superscriptaddress]{revtex4-1}
\usepackage{setspace}
\usepackage{graphicx}
\usepackage{subfigure}
\usepackage{epsfig}
\usepackage{array}
\newcolumntype{M}{ >{\centering\arraybackslash} m{1cm}}
\newcolumntype{P}{ >{\centering\arraybackslash} m{3.85 cm}}
\usepackage{dcolumn}
\usepackage{bm}
\usepackage{amsmath, amsthm, amssymb}
\usepackage[margin=1.0in]{geometry}
\usepackage{url}

\usepackage{xcolor}
\newenvironment{EPedit}{\begingroup\color{red}}{\endgroup}

\newcommand{\beq}{\begin{equation}}
\newcommand{\beqs}{\begin{equation*}}
\newcommand{\eeq}{\end{equation}}
\newcommand{\eeqs}{\end{equation*}}
\newcommand{\bEe}{\begin{EPedit}}
\newcommand{\eEe}{\end{EPedit}}
\newcommand{\rmnum}[1]{\romannumeral#1\relax}
\newcommand{\Rmnum}[1]{\uppercase\expandafter{\romannumeral#1}}

\begin{document}

\title{Instabilities, defects, and defect ordering in an overdamped active nematic}

\author{Elias Putzig}
\email[]{efputzig@brandeis.edu}
\affiliation{Martin Fisher School of Physics, Brandeis University, Waltham, MA 02453, USA}
\author{Gabriel S. Redner}
\email[]{gredner@brandeis.edu}
\affiliation{Martin Fisher School of Physics, Brandeis University, Waltham, MA 02453, USA}
\author{Arvind Baskaran}
\email[]{baskaran@math.uci.edu}
\affiliation{Department of Chemistry and Biochemistry, University of Maryland, College Park, MD 20742, USA}
\author{Aparna Baskaran}
\email[]{aparna@brandeis.edu}
\affiliation{Martin Fisher School of Physics, Brandeis University, Waltham, MA 02453, USA}

\date{\today}

\begin{abstract}
\end{abstract}

\begin{abstract}
We consider a phenomenological continuum theory for an extensile, overdamped active
nematic liquid crystal, applicable in the dense regime. Constructed from general principles, the theory is universal, with parameters independent of any particular microscopic realization. We show that it exhibits a
bend instability similar to that seen in active suspensions, that leads to the proliferation
of defects. We find three distinct nonequilibrium steady states: a defect-ordered nematic in which $+\frac{1}{2}$ disclinations develop polar ordering, an undulating nematic state with no defects, and a turbulent defective nematic. We characterize the phenomenology of these phases and identify the relationship of this theoretical description to experimental realizations and other theoretical models of active nematics.

\end{abstract}

\maketitle

Liquid crystals are anisotropic fluid mesophases that exhibit broken
rotational symmetry and have been extensively studied and extremely useful
for many years \cite{deGennes1993}. The study of topological defects in the
orientational order in these systems, which typically occur under driving, has had a central role in developing our understanding of the material properties of these systems \cite{Chaikin2000, Laverntovich2003}. Active liquid crystals are anisotropic fluid phases that are driven at the scale of the microscopic nematogen. These microscale internal forces give rise to spontaneous defect nucleation, and novel defect dynamics \cite{Sanchez2013, Keber2014, Decamp2015}.

An experimental realization of a two-dimensional active nematic with rich phenomenology is a system of cytoskeletal filaments driven by motor proteins \cite{Sanchez2013, Keber2014, Decamp2015} confined to a fluid interface.  This novel nonequilibrium system has been shown to have transient self-propelled defects and emergent defect ordering and has triggered much theoretical effort to understand its dynamics \cite{Giomi2013, Thampi2013, Blow2014, Giomi2014, Thampi2014a, Thampi2014, Thampi2014b, Gao2015, Shi2013, Decamp2015, Saintillan2007, Saintillan2008, Doostmohammadi2015}.
One outcome of the theories is to show how fluid mediation, in the form of active and passive backflow, can lead to the formation and propulsion of defects.
However, defects also arise in active nematic systems in which fluid mediation plays little or no part, such as vibrated monolayers of granular rods \cite{Narayan2007},
epithelial cell monolayers \cite{Kemkemer2000},
and elongated fibroblasts \cite{Duclos2014}.

\begin{figure}[h!]
  \includegraphics[width=\columnwidth]{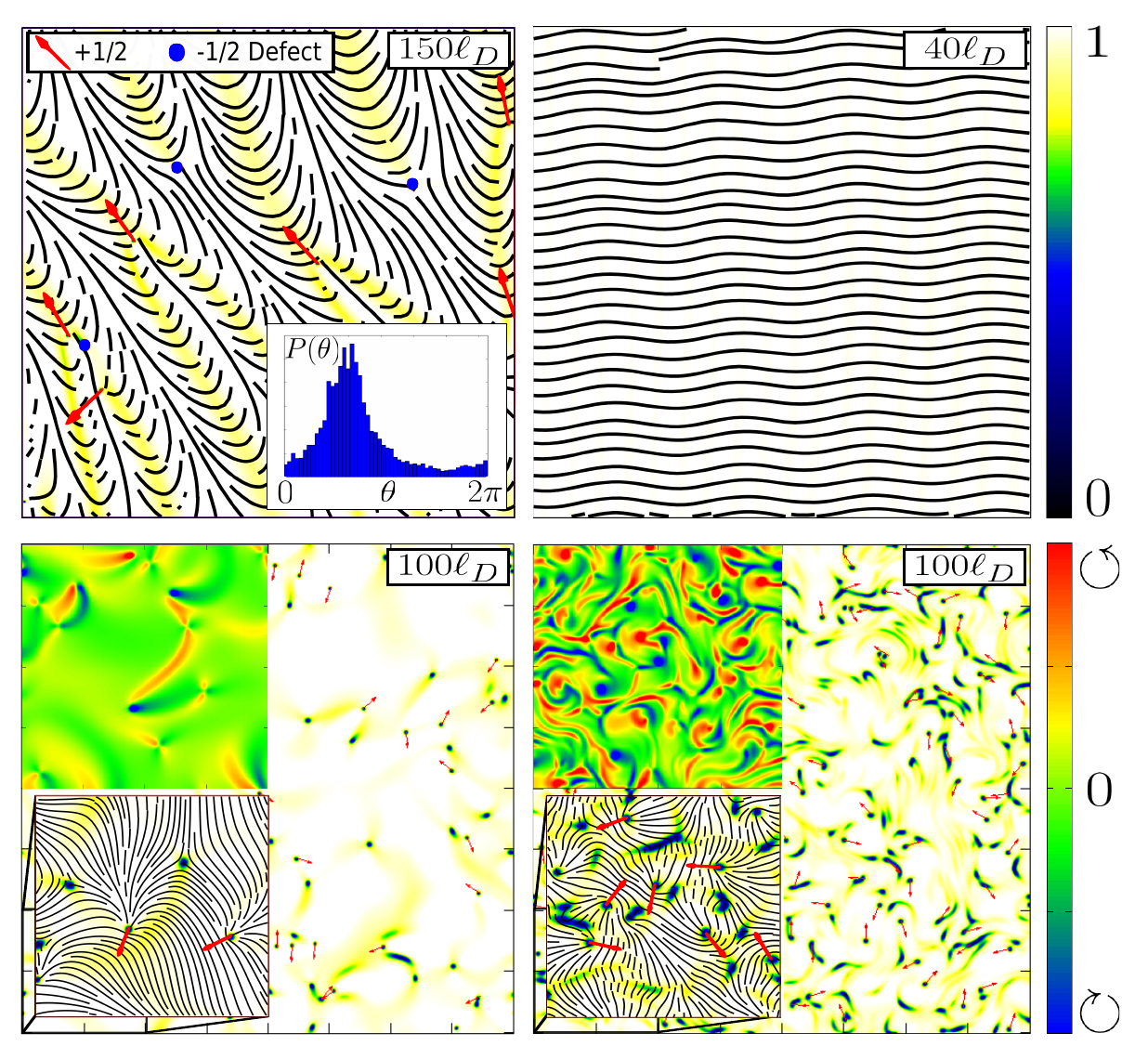}
  \caption{ (color online)  Heatmaps of the degree of order ($S/S_0$, colorbar on the upper-right) showing the nonequilibrium steady-states.  Top Left Panel: The defect-ordered state, with a histogram inset showing the sharp polar ordering in the orientation of $+\frac{1}{2}$ defects. The lines, showing the direction of order, highlight the extended trails left by the motion of these defects. Top Right Panel: The undulating nematic state is highly ordered ($S\simeq S_0$) but the direction of order undulates. Bottom Panels: Defective states at low activity (left, $\psi<1$) and high activity (right, $\psi>1$), with $2\times$-magnified regions (lower left), and heatmaps of the vorticity (in the upper left, colorbar on the lower-right). Scale bars are on the top right. 
}
  \label{States}
\end{figure}

In this work we develop a phenomenological continuum theory that describes the dynamics of an overdamped active nematic, and is applicable to all systems in this symmetry class in two dimensions. The nonequilibrium steady-states that we find are (\rmnum{1}) a defect-ordered nematic that exhibits emergent polar ordering of $+\frac{1}{2}$ defects,  (\rmnum{2}) a defect-free undulating nematic, and (\rmnum{3}) a defective, turbulent nematic (see Fig. \ref{States}). Further, we identify the relationship of existing theories to this framework.

\textbf{Theoretical Framework}: An equilibrium nematic is described by the well-known Landau-de Gennes free energy that is a functional of the density $\rho(\vec{r},t)$ and the nematic order tensor, ${\bf Q}(\vec{r},t)$, associated with rotational symmetry breaking \cite{deGennes1993}. Its dynamics is given by gradient descent on this free energy landscape: `Model A' dynamics for the director,
$\partial_t {\bf Q}(\vec{r})=-\gamma^{-1}\frac{\delta F}{\delta {\bf Q}(\vec{r})} $,
and `Model B' dynamics for the density,
$\partial_t\rho(\vec{r})=-\gamma'^{-1}\nabla^2\frac{\delta F}{\delta \rho(\vec{r})}$,
 as density is conserved. If this system is in an imposed flow then the director is convected
and rotated by that flow, and the dynamical equation becomes
\begin{gather*}
(\partial_{t}+\vec{u}\cdot\vec{\nabla}){\bf Q}=-\gamma^{-1}\frac{\delta F}{\delta {\bf Q}} +{\bf Q}{\bf \Omega} -{\bf \Omega}{\bf Q}\\
+\lambda {\bf E}+\lambda({\bf QE}+ {\bf EQ})^{\mathcal{T}}
\end{gather*}
where ${\bf \Omega}$ is the vorticity tensor, ${\bf E}$ is the
strain-rate tensor associated with the flow, $\lambda$ is the flow-alignment
parameter, and $({\bf A})^{\mathcal{T}}$ denotes the traceless version of ${\bf A}$ 
(e.g. ${\bf A}-\frac{1}{2}Tr({\bf A}){\bf I}$).

In the case of an active liquid crystal, internal stresses due to
the forces exerted by the particles themselves give rise to self-generated
flows. We \emph{postulate} that an active liquid crystal undergoes gradient descent
dynamics in the local rest frame of this \emph{self-generated flow arising from the activity}.
The active stress has the same symmetry as the nematic order ($\sigma\propto{\bf Q}$) and hence the self-generated flow will be proportional to the force density:
 $\rho\vec{u}=\vec{F}/\xi_{T}\equiv-\lambda_{C}\vec{\nabla {\bf Q}}$.
Similarly the vorticity will be proportional to the torque density:
$\rho\vec{\omega}=\vec{\tau}/\xi_{R}\equiv\lambda_{R}(\vec{\nabla}\times\vec{\nabla {\bf Q}})$, where $\xi$'s denote friction coefficients. Using the form of the flow and vorticity
arising from the activity, the dynamical equations
for an active nematic take the form

\begin{gather}
\partial_{t}{\bf Q} =-\gamma^{-1}\frac{\delta F}{\delta {\bf Q}}
+\frac{\lambda_{C}}{\rho}[\vec{\nabla {\bf Q}}]\cdot\vec{\nabla}{\bf Q}
-\frac{\lambda_{S}}{2\rho}{\bf Q} (\boldsymbol{\nabla \nabla}:{\bf Q}) \nonumber \\
- \frac{\lambda_{R}}{\rho}\big( {\bf Q}\cdot \boldsymbol{\nabla \nabla} \cdot {\bf Q} -\vec{\nabla}(\vec{\nabla {\bf Q}}) \cdot {\bf Q}\big)^{\mathcal{S}}  \label{QDynam}
\end{gather}
where the coefficients $\lambda_{C}$, $\lambda_{R}$, and $\lambda_{S}$ control the strength
of active convection, active torque, and flow-alignment respectively, ${\bf A:B}=A_{ij}B_{ij}$, and ${\mathcal{S}}$ denotes symmetrization (e.g. $\big({\bf A}\big)^{\mathcal{S}}=\frac{1}{2}({\bf A + A^T})$). Further, we take the dynamics of the density to be of the form
\beq \label{rhoDynam}
\partial_t\rho=D\nabla^2\rho+D_Q\boldsymbol{\nabla \nabla}:{\bf Q}
\eeq
which incorporates the self-generated flow through the curvature induced density flux \cite{Simha2002, Ramaswamy2003}, controlled by $D_Q$.

Equations (\ref{QDynam}) and (\ref{rhoDynam}) are the dynamical equations of an overdamped active nematic which will be studied in this work. Before we proceed with the analysis of our theory, we make the following observations in order to place this model in the context of other theories in the literature of active liquid crystals:

\noindent(1) Suppose our active fluid had polar symmetry, the self generated flow will be proportional to the polar order parameter, $\rho\vec{u}=\vec{F}/\xi_T \equiv \lambda\vec{P}$. Postulating a gradient descent dynamics, as above, for an active polar fluid results in the same dynamics as described in the seminal works of Toner and Tu \cite{Toner1995, Toner1998}. In this sense, this theory is a generalization of their approach to active nematic systems.

\noindent(2) Existing theories of active nematics consider a coupled set of equations of the nematic order parameter and the active induced flow which arises from a Stokes equation (such as \cite{Giomi2013, Thampi2013, Blow2014, Giomi2014, Thampi2014a, Thampi2014, Thampi2014b, Doostmohammadi2015}). Suppose there exists a screening mechanism such as confinement to 2D, then eliminating the flow field in terms of the active stress yields Eqs (\ref{QDynam}) and (\ref{rhoDynam}) above, but with
$\lambda_C=\lambda_R$  i.e., a Galilean invariant version of our theory \cite{Doostmohammadi2015, Pismen2013a, Communication2015, MarchMeeting2015}. This is an important distinction of our work. In overdamped systems Galilean invariance is broken by the medium through which our particles move (as first pointed out in \cite{Toner1995}) and this is accounted for in our theory.

\noindent(3) Theoretical work on active nematics was pioneered in \cite{Simha2002, Ramaswamy2003}, and have subsequently been shown to have giant number fluctuations, phase separation, and band formation near the critical density \cite{Mishra2006, Chate2006, Narayan2007, Baskaran2008, Baskaran2008a, Mishra2010, Peruani2011, Peshkov2012, Peshkov2012a, Baskaran2012, Bertin2013, Ngo2013, Putzig2014}. Our Eqs (\ref{QDynam}) and (\ref{rhoDynam}) reduce to this description when $\lambda_R=\lambda_C=\lambda_S=0$. Hence our theory can be considered a generalization of these previous works.

\textbf{Parameters of the theory}: The equilibrium contributions to the dynamics of the order parameter take the form
\begin{gather*}
    [\partial_t Q_{ij}]_{Eq}
  = D_r[\alpha-\beta Tr{\bf Q}^{2}] Q_{ij}
+ 2D_E \nabla^{2} Q_{ij}  \\
+ \frac{D_{\delta}}{\rho}  \big( 2\partial_k(Q_{kl}\partial_lQ_{ij})
- ([\partial_{i}Q_{kl}]\partial_{j}Q_{kl})^{\mathcal{T}} \big) \\
+ D_{\rho}(\partial_{i}\partial_{j})^{\mathcal{T}}\rho
- K \nabla^4 Q_{i j}
\end{gather*}
where $\alpha=(\rho-1)$ and $\beta=\frac{1}{\rho^{2}}(\rho+1)$. $D_{r}$ is the rotational diffusion constant and $D_{\rho}$ is a kinetic term also seen in prior works \cite{Baskaran2008, Baskaran2008a,Peshkov2012a}. There are two elastic terms; $D_E$ is a mean elasticity, and $D_\delta$ is a differential elasticity, measuring the difference between bend and splay energies. Finally, a fourth-order gradient term (with coefficient $K$) is included in order to ensure smoothness and numerical stability. The relevant parameters for the phenomenology discussed are: the active force and torque ($\lambda_{C,R}$) and the differential elastic constant $D_\delta$. In the following, we non-dimensionalize our equations by setting our time scale to be the rotational diffusion time, $\frac{1}{D_r}$, and our length-scale to be the diffusion length, $\ell_D\equiv \sqrt{D/D_r}$. In all of the subsequent sections we will work in these dimensionless variables.

\textbf{Instabilities of the Nematic State}: In the homogeneous limit, Eqs. (\ref{QDynam}) and (\ref{rhoDynam}) admit a uni-axial nematic solution with average density $\rho_0>1$, and the order parameter ${\bf Q}=\rho_0S_0(\boldsymbol{\hat{x}\hat{x}}-\frac{1}{2}{\bf I})$ with  $S_0=\sqrt{\frac{2(\rho_0-1)}{\rho_0+1}}$. Let us consider spatial fluctuations about this state.

Fluctuations perpendicular to the director cause an instability when
$ D_{Q}>\sqrt{2(\rho _{0}^{2}-1)}\Big(\frac{\rho _{0}+1}{\rho_{0}^{2}+\rho _{0}-1}\Big) $.
This instability causes phase separation into bands of dense ordered regions coexisting with dilute disordered regions when the material is near the critical density ($\rho_c=1$). This has been discussed in previous work by us \cite{Putzig2014} and others \cite{Peruani2011, Peshkov2012, Peshkov2012a, Bertin2013, Ngo2013}.   The nonlinear active terms ($\lambda_C$, $\lambda_R$ and $\lambda_S$) do not alter the instability, nor do they significantly alter the phenomenology discussed in previous work.

Of primary interest here is a bend instability in which the direction of ordering is destabilized by fluctuations parallel to the director.  This occurs when a `bend instability parameter' $\psi>1$, where
$\psi = \frac{\lambda_R-2D_\delta}{D_E\cdot \Lambda (\rho_0)}$, and
 $\Lambda (\rho_0)=4\sqrt{\frac{\rho_0+1}{2(\rho_0-1)}}$. This instability parameter reflects a competition between the active torque $\lambda_R$ and the differential elastic constant
$D_\delta$, which is positive if the energetic cost to bend distortions of the director is greater than that of splay. The bend instability is the primary mechanism for defect generation and formation of inhomogeneous steady states in this active system.

\begin{figure}[t]
  \includegraphics[width=\columnwidth]{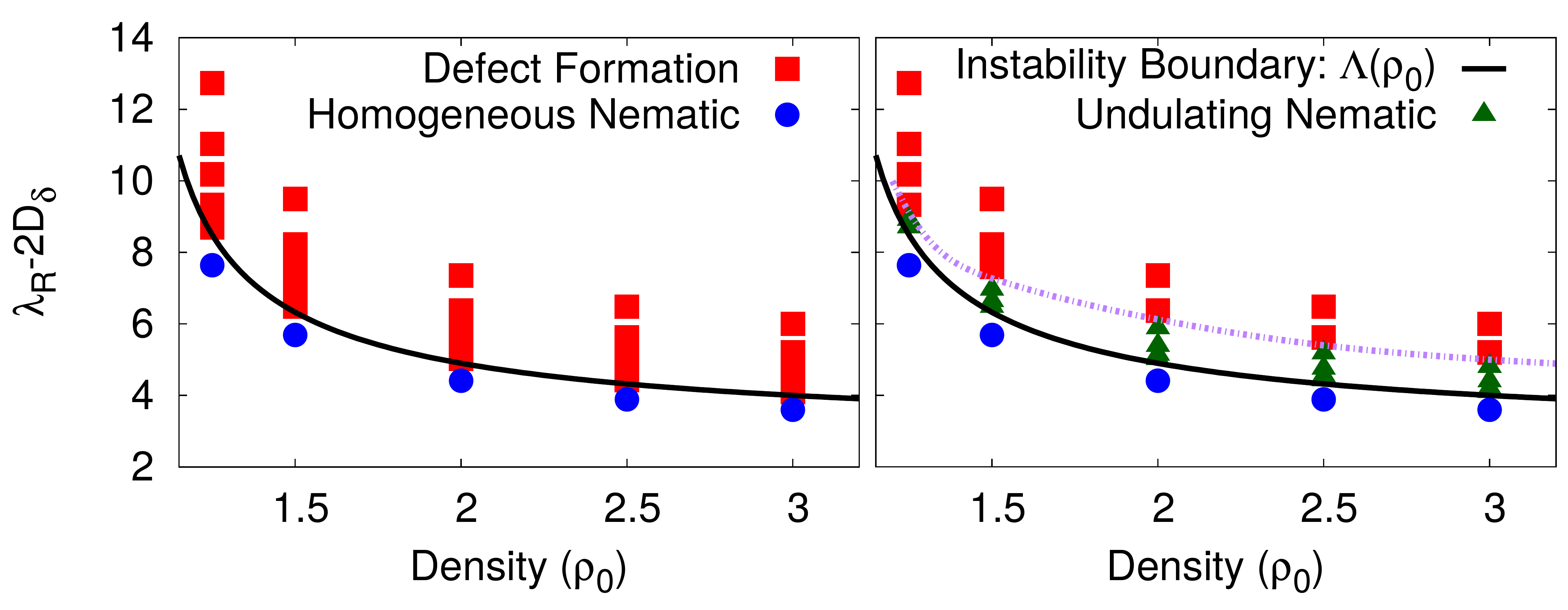}
  \caption{ (color online) Plots showing the end state which forms from nematic initial conditions. Left: At $D_\delta =1.0$, the homogeneous nematic state transitions into a defective nematic. Right: At $D_\delta =-0.50$ we find an undulating state at intermediate activities.  }
  \label{PlotBI}
\end{figure}

In order to elucidate the consequence of this bend instability, we numerically explored the dynamics using a semi-implicit finite difference method, with periodic boundary conditions. Integrating from nematic initial conditions with small amplitude Gaussian noise, two states were found above the bend instability ($\psi>1$). These were (\Rmnum{1}) an undulating nematic state (see Fig. \ref{PlotBI}) where the system is strongly ordered but the director undulates along the broken symmetry direction (see Fig. \ref{States}), and (\Rmnum{2}) a turbulent state (see Fig. \ref{PlotBI},\ref{States}) in which charge $\pm\frac{1}{2}$ disclinations continually form and annihilate and the $+\frac{1}{2}$ defects are self-propelled as seen in \cite{Sanchez2013, Shi2013, Thampi2013, Keber2014, Blow2014, Gao2015, Giomi2014, Giomi2013, Blow2014, Giomi2014, Thampi2014a, Thampi2014, Thampi2014b,  Decamp2015}. The undulating nematic state was present when active convection $\lambda_C>1.0$ and when $D_\delta\preceq 0$. In other regions of parameter space the system transitioned directly into the defective nematic state.
\footnote{Details of the this transition, as well as the additional discussion concerning the equations and linear stability analysis, can be found in the supplement.}.

The above analysis focused on the instabilities of the homogeneous nematic state. Next we consider isotropic initial conditions and present the results for different values of the parameters that control the bend instability ($\lambda_R$, $D_\delta$, and $\rho_0$) while keeping the other parameters fixed ($\lambda_C=D_\rho=D_E=D=1.0$, $K=0.50$, $\lambda_S =0$, and $D_Q=0.10$) unless otherwise specified.

\begin{figure}
  \includegraphics[width=\columnwidth]{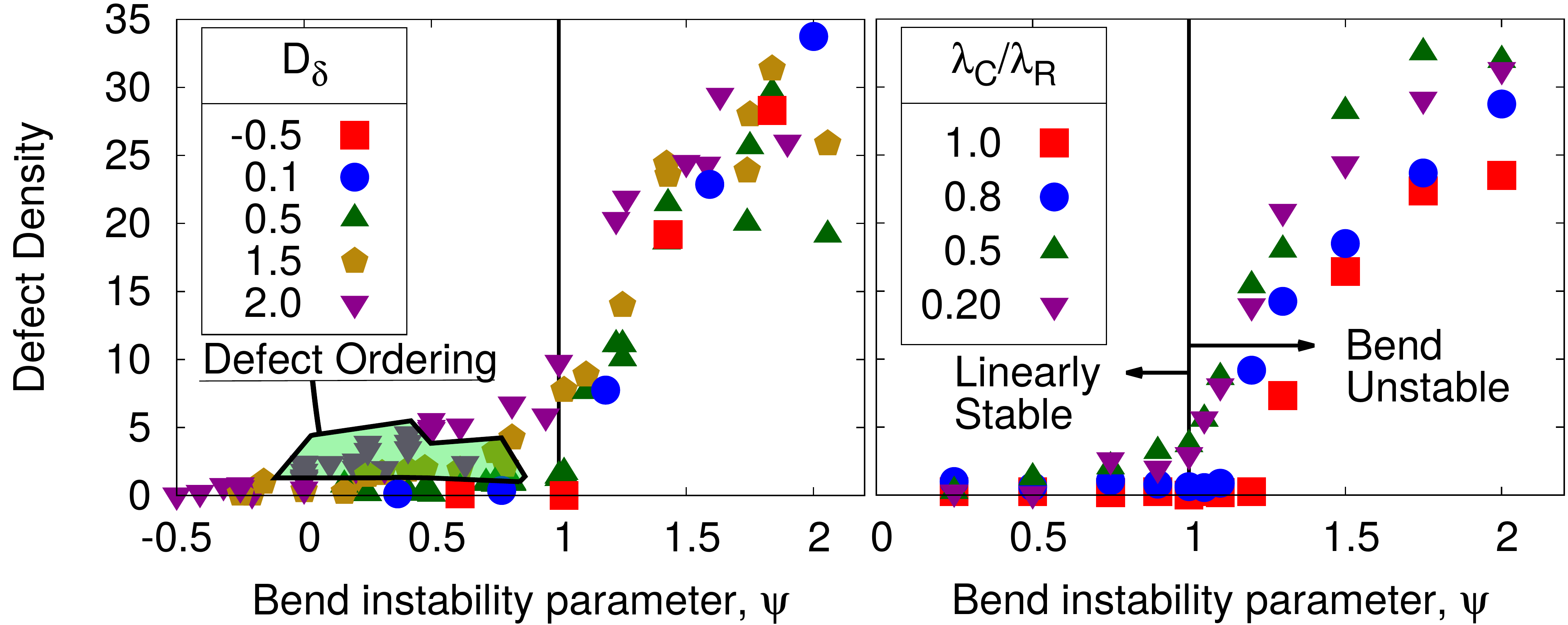}
  \caption{ (color online)  Plots of the defect density $n$ (defects per $(100\ell_D)^2$) as a function of $\psi$. Left: Curves with fixed $\lambda_C=1.0$, a range of $D_\delta$ and $\rho_0$ (scaled as  $n/(\rho_0-1)$). Defect density is greater for larger $D_\delta$ when $\psi<1$, increases sharply when $\psi$ crosses $1$, then saturates. The shaded region indicates where there was statistically significant polar ordering of $+\frac{1}{2}$ defects. Right: Fixed $D_\delta=1.0$, for a range of $\lambda_C/\lambda_R$, shows that when the active convection is comparable to active torque, defects vanish near the bend instability. 
}
  \label{DefectComp}
\end{figure}

\textbf{Defect-ordered state}: This nonequilibrium steady state occurs when defects form below the bend instability ($\psi<1$). This occurs when the strength of active torque is dominant over the strength of active convection (see Fig. \ref{DefectComp}). The properties of this state are as follows: (\rmnum{1}) Defects are point-like and the background is a well-ordered nematic (see Fig. \ref{States}). (\rmnum{2}) Defining the orientation of $+\frac{1}{2}$ defects to be opposite the ``comet tail' (along the direction of propulsion), we find that these defects exhibit \emph{significant polar ordering}. (\rmnum{3}) The degree of polar ordering decreases as $\psi$ increases. (\rmnum{4}) The high degree of polar ordering corresponds with long splay distortions which are left by $+\frac{1}{2}$ defects as they travel (see Fig. \ref{States}). Other $+\frac{1}{2}$ defects tend to reorient rather than cross this distortion-trail, leading to long parallel structures which are visible in the states with a large degree of polar ordering.

\begin{figure}
  \includegraphics[width= \columnwidth]{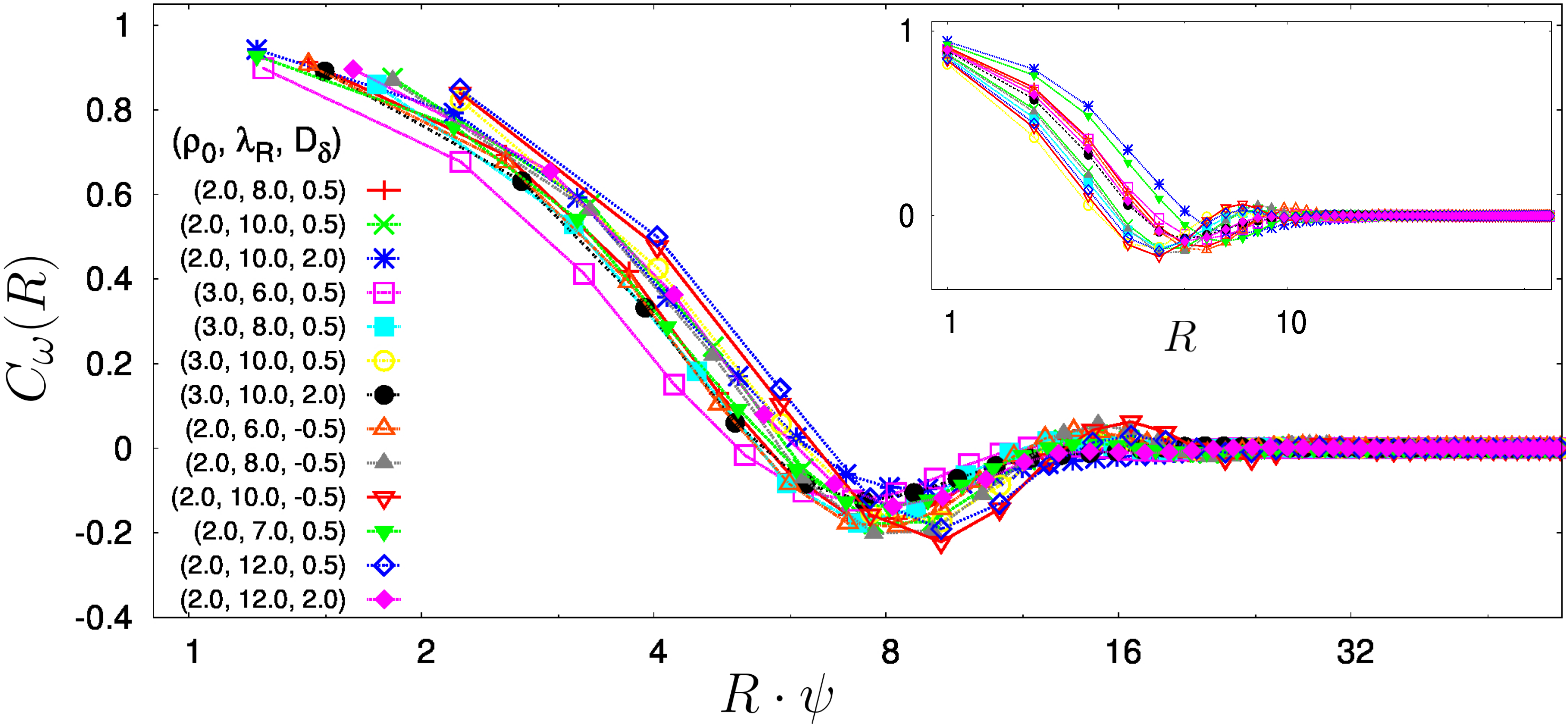}
  \caption{ (color online)  Log-linear plot of vorticity correlation functions, $C_\omega(R)$ for fixed parameters ($\lambda_R$, $D_\delta$ and $\rho_0$) above the bend instability ($\psi>1.10$). The length is scaled by $\psi$, which gives a data collapse for a large range of parameters.
}
\label{VortS}
\end{figure}

\textbf{Turbulent nematic state}: The turbulent, defective nematic state occurs just above the bend instability and has the following properties:
(\rmnum{1}) Defect density and vorticity increase sharply for $\psi \sim 1$.
(\rmnum{2}) There is a saturation, or even a decrease, in the defect density (see Fig. \ref{DefectComp}) as the defects, which were point-like and circular near $\psi=1$, become spatially extended and the average degree of ordering decreases.
(\rmnum{3}) The vorticity correlation function
$C_\omega (R)=\langle \omega(0) \omega(R)\rangle /  \langle \omega^2 \rangle$, scales with the bend instability parameter $\psi$ (see Fig \ref{VortS}), which is linear in the strength of the active torque $\lambda_R$. This differs from what was found in a fluid-mediated active nematic theory \cite{Thampi2014} where vorticity scaled to the $1/4$th power of the strength of the activity. (\rmnum{4}) Assuming that the length scale for defect separation $\ell_d$ scales with the vorticity ($\ell_d\propto \frac{1}{\psi}$) would lead to a prediction that defect density scales as $\psi^2$. This seems compatible with the trend seen near the critical value of the bend instability parameter, but the range is not large enough for a conclusive comparison.

\textbf{Discussion}: We have introduced a universal model of an overdamped active nematic in which activity enters through self-induced flows. This theory encompasses existing work on active nematics as special cases. The active torque from self-generated flows gives rise to a bend instability which, in turn, leads to defect formation.  We have identified three nonequilibrium steady states admitted by this theory. The first is a defect-ordered nematic state where polar ordering of $+1/2$ disclinations emerges from the underlying apolar theory. The second is an undulating nematic state which is reminiscent of the ``walls'' of distortion in the order parameter seen before the onset of defective states \cite{Thampi2014,Giomi2014}, or the distortion of the director which happens during relaxation oscillations \cite{Giomi2011, Giomi2012} in active nematic suspensions. Finally we find a turbulent nematic state similar to that which occurs in theories of active nematic suspension \cite{Giomi2013, Thampi2013, Blow2014, Giomi2014, Thampi2014a, Thampi2014, Thampi2014b}.

The appearance of a defect-ordered state is of particular interest, as defect-ordering was recently discovered in layers of active cytoskeletal filaments and in simulations of that system \cite{Decamp2015}. Our theory provides robust predictions about when defect ordering will be found. The ordering occurs below the bend instability, i.e., $\psi<1$, and only for $D_\delta>0$, which indicates that the energetic penalty to bend distortions must be greater than that of splay, and the strength of the active torque must be greater than that of the active convection. This result implies that an orientationally ordered fluid phase of defects may not occur in theories which have Galilean invariance ($\lambda_C=\lambda_R$).

In \cite{Decamp2015}, defect ordering in the experiment had nematic symmetry, while simulations displayed the same polar symmetry that we observe. This may be due to the use of periodic boundary conditions in both the simulations in \cite{Decamp2015} and our work, or the nematic ordering could be a subtle effect of fluid-mediated interactions in the experiment. However, the strong similarities between this theory and the experimental system, such as the sharp splay distortions left in the wake of traveling $+\frac{1}{2}$ defects leads us to believe that the parameter space in which this system operates is $\psi<1$, $\lambda_C<\lambda_R$ and $D_\delta>0$. Further work is needed to substantiate this in detail.

The theory that we have described contains a rich and diverse phenomenology which overlaps with other theories, simulations, and experiments of active nematics. It is generic and universal in that it is independent of any particular microscopic model and only relies on the nematic having local interactions.

\begin{acknowledgments}
We thank Mike Hagan, Zvonimir Dogic, and Stephen DeCamp for sharing information and ideas. EFP and AB acknowledge support from NSF-DMR-1149266, the Brandeis-MRSEC through NSF DMR-0820492 and NSF MRSEC-1206146, NSF PHY11-25915 through KITP,  and the HPC cluster at Brandeis for computing time. EFP also acknowledges support through NIH-5T32EB009419 and IGERT DGE-1068620.

\end{acknowledgments}

\bibliography{References}

\begin{thebibliography}{43}%
\makeatletter
\providecommand \@ifxundefined [1]{%
 \@ifx{#1\undefined}
}%
\providecommand \@ifnum [1]{%
 \ifnum #1\expandafter \@firstoftwo
 \else \expandafter \@secondoftwo
 \fi
}%
\providecommand \@ifx [1]{%
 \ifx #1\expandafter \@firstoftwo
 \else \expandafter \@secondoftwo
 \fi
}%
\providecommand \natexlab [1]{#1}%
\providecommand \enquote  [1]{``#1''}%
\providecommand \bibnamefont  [1]{#1}%
\providecommand \bibfnamefont [1]{#1}%
\providecommand \citenamefont [1]{#1}%
\providecommand \href@noop [0]{\@secondoftwo}%
\providecommand \href [0]{\begingroup \@sanitize@url \@href}%
\providecommand \@href[1]{\@@startlink{#1}\@@href}%
\providecommand \@@href[1]{\endgroup#1\@@endlink}%
\providecommand \@sanitize@url [0]{\catcode `\\12\catcode `\$12\catcode
  `\&12\catcode `\#12\catcode `\^12\catcode `\_12\catcode `\%12\relax}%
\providecommand \@@startlink[1]{}%
\providecommand \@@endlink[0]{}%
\providecommand \url  [0]{\begingroup\@sanitize@url \@url }%
\providecommand \@url [1]{\endgroup\@href {#1}{\urlprefix }}%
\providecommand \urlprefix  [0]{URL }%
\providecommand \Eprint [0]{\href }%
\providecommand \doibase [0]{http://dx.doi.org/}%
\providecommand \selectlanguage [0]{\@gobble}%
\providecommand \bibinfo  [0]{\@secondoftwo}%
\providecommand \bibfield  [0]{\@secondoftwo}%
\providecommand \translation [1]{[#1]}%
\providecommand \BibitemOpen [0]{}%
\providecommand \bibitemStop [0]{}%
\providecommand \bibitemNoStop [0]{.\EOS\space}%
\providecommand \EOS [0]{\spacefactor3000\relax}%
\providecommand \BibitemShut  [1]{\csname bibitem#1\endcsname}%
\let\auto@bib@innerbib\@empty
\bibitem [{\citenamefont {De~Gennes}\ and\ \citenamefont
  {Prost}(1993)}]{deGennes1993}%
  \BibitemOpen
  \bibfield  {author} {\bibinfo {author} {\bibfnamefont {P.-G.}\ \bibnamefont
  {De~Gennes}}\ and\ \bibinfo {author} {\bibfnamefont {J.}~\bibnamefont
  {Prost}},\ }\href@noop {} {\emph {\bibinfo {title} {The physics of liquid
  crystals}}},\ Vol.~\bibinfo {volume} {23}\ (\bibinfo  {publisher} {Clarendon
  press Oxford},\ \bibinfo {year} {1993})\BibitemShut {NoStop}%
\bibitem [{\citenamefont {Chaikin}\ and\ \citenamefont
  {Lubensky}(2000)}]{Chaikin2000}%
  \BibitemOpen
  \bibfield  {author} {\bibinfo {author} {\bibfnamefont {P.~M.}\ \bibnamefont
  {Chaikin}}\ and\ \bibinfo {author} {\bibfnamefont {T.~C.}\ \bibnamefont
  {Lubensky}},\ }\href@noop {} {\emph {\bibinfo {title} {Principles of
  condensed matter physics}}},\ Vol.~\bibinfo {volume} {1}\ (\bibinfo
  {publisher} {Cambridge Univ Press},\ \bibinfo {year} {2000})\BibitemShut
  {NoStop}%
\bibitem [{\citenamefont {Laverntovich}(2003)}]{Laverntovich2003}%
  \BibitemOpen
  \bibfield  {author} {\bibinfo {author} {\bibfnamefont {O.~D.}\ \bibnamefont
  {Laverntovich}},\ }\href@noop {} {\emph {\bibinfo {title} {Soft matter
  physics: an introduction}}}\ (\bibinfo  {publisher} {Springer Science \&
  Business Media},\ \bibinfo {year} {2003})\BibitemShut {NoStop}%
\bibitem [{\citenamefont {Sanchez}\ \emph {et~al.}(2013)\citenamefont
  {Sanchez}, \citenamefont {Chen}, \citenamefont {DeCamp}, \citenamefont
  {Heymann},\ and\ \citenamefont {Dogic}}]{Sanchez2013}%
  \BibitemOpen
  \bibfield  {author} {\bibinfo {author} {\bibfnamefont {T.}~\bibnamefont
  {Sanchez}}, \bibinfo {author} {\bibfnamefont {D.~T.~N.}\ \bibnamefont
  {Chen}}, \bibinfo {author} {\bibfnamefont {S.~J.}\ \bibnamefont {DeCamp}},
  \bibinfo {author} {\bibfnamefont {M.}~\bibnamefont {Heymann}}, \ and\
  \bibinfo {author} {\bibfnamefont {Z.}~\bibnamefont {Dogic}},\ }\href
  {\doibase 10.1038/nature11591} {\bibfield  {journal} {\bibinfo  {journal}
  {Nature}\ ,\ \bibinfo {pages} {24}} (\bibinfo {year} {2013})},\ \Eprint
  {http://arxiv.org/abs/1301.1122} {1301.1122} \BibitemShut {NoStop}%
\bibitem [{\citenamefont {Keber}\ \emph {et~al.}(2014)\citenamefont {Keber},
  \citenamefont {Loiseau}, \citenamefont {Sanchez}, \citenamefont {DeCamp},
  \citenamefont {Giomi}, \citenamefont {Bowick}, \citenamefont {Marchetti},
  \citenamefont {Dogic},\ and\ \citenamefont {Bausch}}]{Keber2014}%
  \BibitemOpen
  \bibfield  {author} {\bibinfo {author} {\bibfnamefont {F.~C.}\ \bibnamefont
  {Keber}}, \bibinfo {author} {\bibfnamefont {E.}~\bibnamefont {Loiseau}},
  \bibinfo {author} {\bibfnamefont {T.}~\bibnamefont {Sanchez}}, \bibinfo
  {author} {\bibfnamefont {S.~J.}\ \bibnamefont {DeCamp}}, \bibinfo {author}
  {\bibfnamefont {L.}~\bibnamefont {Giomi}}, \bibinfo {author} {\bibfnamefont
  {M.~J.}\ \bibnamefont {Bowick}}, \bibinfo {author} {\bibfnamefont {M.~C.}\
  \bibnamefont {Marchetti}}, \bibinfo {author} {\bibfnamefont {Z.}~\bibnamefont
  {Dogic}}, \ and\ \bibinfo {author} {\bibfnamefont {A.~R.}\ \bibnamefont
  {Bausch}},\ }\href {\doibase 10.1126/science.1254784} {\bibfield  {journal}
  {\bibinfo  {journal} {Science}\ }\textbf {\bibinfo {volume} {345}},\ \bibinfo
  {pages} {1} (\bibinfo {year} {2014})}\BibitemShut {NoStop}%
\bibitem [{\citenamefont {Decamp}\ \emph {et~al.}(2015)\citenamefont {Decamp},
  \citenamefont {Redner}, \citenamefont {Baskaran}, \citenamefont {Hagan},\
  and\ \citenamefont {Dogic}}]{Decamp2015}%
  \BibitemOpen
  \bibfield  {author} {\bibinfo {author} {\bibfnamefont {S.~J.}\ \bibnamefont
  {Decamp}}, \bibinfo {author} {\bibfnamefont {G.~S.}\ \bibnamefont {Redner}},
  \bibinfo {author} {\bibfnamefont {A.}~\bibnamefont {Baskaran}}, \bibinfo
  {author} {\bibfnamefont {M.~F.}\ \bibnamefont {Hagan}}, \ and\ \bibinfo
  {author} {\bibfnamefont {Z.}~\bibnamefont {Dogic}},\ }\href@noop {}
  {\bibfield  {journal} {\bibinfo  {journal} {arXiv.org}\ ,\ \bibinfo {pages}
  {1}} (\bibinfo {year} {2015})},\ \Eprint
  {http://arxiv.org/abs/arXiv:1501.06228v2} {arXiv:arXiv:1501.06228v2}
  \BibitemShut {NoStop}%
\bibitem [{\citenamefont {Giomi}\ \emph {et~al.}(2013)\citenamefont {Giomi},
  \citenamefont {Bowick}, \citenamefont {Ma},\ and\ \citenamefont
  {Marchetti}}]{Giomi2013}%
  \BibitemOpen
  \bibfield  {author} {\bibinfo {author} {\bibfnamefont {L.}~\bibnamefont
  {Giomi}}, \bibinfo {author} {\bibfnamefont {M.~J.}\ \bibnamefont {Bowick}},
  \bibinfo {author} {\bibfnamefont {X.}~\bibnamefont {Ma}}, \ and\ \bibinfo
  {author} {\bibfnamefont {M.~C.}\ \bibnamefont {Marchetti}},\ }\href {\doibase
  10.1103/PhysRevLett.110.228101} {\bibfield  {journal} {\bibinfo  {journal}
  {Phys. Rev. Lett.}\ }\textbf {\bibinfo {volume} {110}},\ \bibinfo {pages}
  {228101} (\bibinfo {year} {2013})}\BibitemShut {NoStop}%
\bibitem [{\citenamefont {Thampi}\ \emph {et~al.}(2013)\citenamefont {Thampi},
  \citenamefont {Golestanian},\ and\ \citenamefont {Yeomans}}]{Thampi2013}%
  \BibitemOpen
  \bibfield  {author} {\bibinfo {author} {\bibfnamefont {S.~P.}\ \bibnamefont
  {Thampi}}, \bibinfo {author} {\bibfnamefont {R.}~\bibnamefont {Golestanian}},
  \ and\ \bibinfo {author} {\bibfnamefont {J.~M.}\ \bibnamefont {Yeomans}},\
  }\href {\doibase 10.1103/PhysRevLett.111.118101} {\bibfield  {journal}
  {\bibinfo  {journal} {Phys. Rev. Lett}\ }\textbf {\bibinfo {volume} {111}},\
  \bibinfo {pages} {118101} (\bibinfo {year} {2013})}\BibitemShut {NoStop}%
\bibitem [{\citenamefont {Blow}\ \emph {et~al.}(2014)\citenamefont {Blow},
  \citenamefont {Thampi},\ and\ \citenamefont {Yeomans}}]{Blow2014}%
  \BibitemOpen
  \bibfield  {author} {\bibinfo {author} {\bibfnamefont {M.~L.}\ \bibnamefont
  {Blow}}, \bibinfo {author} {\bibfnamefont {S.~P.}\ \bibnamefont {Thampi}}, \
  and\ \bibinfo {author} {\bibfnamefont {J.~M.}\ \bibnamefont {Yeomans}},\
  }\href {\doibase 10.1103/PhysRevLett.113.248303} {\bibfield  {journal}
  {\bibinfo  {journal} {Phys. Rev. Lett.}\ }\textbf {\bibinfo {volume} {113}},\
  \bibinfo {pages} {248303} (\bibinfo {year} {2014})}\BibitemShut {NoStop}%
\bibitem [{\citenamefont {Giomi}\ \emph {et~al.}(2014)\citenamefont {Giomi},
  \citenamefont {Bowick}, \citenamefont {Mishra}, \citenamefont {Sknepnek},\
  and\ \citenamefont {Marchetti}}]{Giomi2014}%
  \BibitemOpen
  \bibfield  {author} {\bibinfo {author} {\bibfnamefont {L.}~\bibnamefont
  {Giomi}}, \bibinfo {author} {\bibfnamefont {M.~J.}\ \bibnamefont {Bowick}},
  \bibinfo {author} {\bibfnamefont {P.}~\bibnamefont {Mishra}}, \bibinfo
  {author} {\bibfnamefont {R.}~\bibnamefont {Sknepnek}}, \ and\ \bibinfo
  {author} {\bibfnamefont {M.~C.}\ \bibnamefont {Marchetti}},\ }\href@noop {}
  {\bibfield  {journal} {\bibinfo  {journal} {Philosophical Transactions of the
  Royal Society A: Mathematical, Physical and Engineering Sciences}\ }\textbf
  {\bibinfo {volume} {372}},\ \bibinfo {pages} {20130365} (\bibinfo {year}
  {2014})}\BibitemShut {NoStop}%
\bibitem [{\citenamefont {Thampi}\ \emph
  {et~al.}(2014{\natexlab{a}})\citenamefont {Thampi}, \citenamefont
  {Golestanian},\ and\ \citenamefont {Yeomans}}]{Thampi2014a}%
  \BibitemOpen
  \bibfield  {author} {\bibinfo {author} {\bibfnamefont {S.~P.}\ \bibnamefont
  {Thampi}}, \bibinfo {author} {\bibfnamefont {R.}~\bibnamefont {Golestanian}},
  \ and\ \bibinfo {author} {\bibfnamefont {J.~M.}\ \bibnamefont {Yeomans}},\
  }\href {\doibase 10.1103/PhysRevE.90.062307} {\bibfield  {journal} {\bibinfo
  {journal} {Physical Review E}\ }\textbf {\bibinfo {volume} {90}},\ \bibinfo
  {pages} {062307} (\bibinfo {year} {2014}{\natexlab{a}})}\BibitemShut
  {NoStop}%
\bibitem [{\citenamefont {Thampi}\ \emph
  {et~al.}(2014{\natexlab{b}})\citenamefont {Thampi}, \citenamefont
  {Golestanian},\ and\ \citenamefont {Yeomans}}]{Thampi2014}%
  \BibitemOpen
  \bibfield  {author} {\bibinfo {author} {\bibfnamefont {S.~P.}\ \bibnamefont
  {Thampi}}, \bibinfo {author} {\bibfnamefont {R.}~\bibnamefont {Golestanian}},
  \ and\ \bibinfo {author} {\bibfnamefont {J.~M.}\ \bibnamefont {Yeomans}},\
  }\href@noop {} {\bibfield  {journal} {\bibinfo  {journal} {Philosophical
  Transactions of the Royal Society A: Mathematical, Physical and Engineering
  Sciences}\ }\textbf {\bibinfo {volume} {372}},\ \bibinfo {pages} {20130366}
  (\bibinfo {year} {2014}{\natexlab{b}})}\BibitemShut {NoStop}%
\bibitem [{\citenamefont {Thampi}\ \emph
  {et~al.}(2014{\natexlab{c}})\citenamefont {Thampi}, \citenamefont
  {Golestanian},\ and\ \citenamefont {Yeomans}}]{Thampi2014b}%
  \BibitemOpen
  \bibfield  {author} {\bibinfo {author} {\bibfnamefont {S.~P.}\ \bibnamefont
  {Thampi}}, \bibinfo {author} {\bibfnamefont {R.}~\bibnamefont {Golestanian}},
  \ and\ \bibinfo {author} {\bibfnamefont {J.~M.}\ \bibnamefont {Yeomans}},\
  }\href {\doibase 10.1209/0295-5075/105/18001} {\bibfield  {journal} {\bibinfo
   {journal} {EPL (Europhysics Letters)}\ }\textbf {\bibinfo {volume} {105}},\
  \bibinfo {pages} {18001} (\bibinfo {year} {2014}{\natexlab{c}})}\BibitemShut
  {NoStop}%
\bibitem [{\citenamefont {Gao}\ \emph {et~al.}(2015)\citenamefont {Gao},
  \citenamefont {Blackwell}, \citenamefont {Glaser}, \citenamefont
  {Betterton},\ and\ \citenamefont {Shelley}}]{Gao2015}%
  \BibitemOpen
  \bibfield  {author} {\bibinfo {author} {\bibfnamefont {T.}~\bibnamefont
  {Gao}}, \bibinfo {author} {\bibfnamefont {R.}~\bibnamefont {Blackwell}},
  \bibinfo {author} {\bibfnamefont {M.~A.}\ \bibnamefont {Glaser}}, \bibinfo
  {author} {\bibfnamefont {M.}~\bibnamefont {Betterton}}, \ and\ \bibinfo
  {author} {\bibfnamefont {M.~J.}\ \bibnamefont {Shelley}},\ }\href@noop {}
  {\bibfield  {journal} {\bibinfo  {journal} {Physical review letters}\
  }\textbf {\bibinfo {volume} {114}},\ \bibinfo {pages} {048101} (\bibinfo
  {year} {2015})}\BibitemShut {NoStop}%
\bibitem [{\citenamefont {Shi}\ and\ \citenamefont {Ma}(2013)}]{Shi2013}%
  \BibitemOpen
  \bibfield  {author} {\bibinfo {author} {\bibfnamefont {X.-q.}\ \bibnamefont
  {Shi}}\ and\ \bibinfo {author} {\bibfnamefont {Y.-q.}\ \bibnamefont {Ma}},\
  }\href {\doibase 10.1038/ncomms4013} {\bibfield  {journal} {\bibinfo
  {journal} {Nat. Commun.}\ }\textbf {\bibinfo {volume} {4}},\ \bibinfo {pages}
  {3013} (\bibinfo {year} {2013})}\BibitemShut {NoStop}%
\bibitem [{\citenamefont {Saintillan}\ and\ \citenamefont
  {Shelley}(2007)}]{Saintillan2007}%
  \BibitemOpen
  \bibfield  {author} {\bibinfo {author} {\bibfnamefont {D.}~\bibnamefont
  {Saintillan}}\ and\ \bibinfo {author} {\bibfnamefont {M.~J.}\ \bibnamefont
  {Shelley}},\ }\href {\doibase 10.1103/PhysRevLett.99.058102} {\bibfield
  {journal} {\bibinfo  {journal} {Physical Review Letters}\ }\textbf {\bibinfo
  {volume} {99}},\ \bibinfo {pages} {1} (\bibinfo {year} {2007})}\BibitemShut
  {NoStop}%
\bibitem [{\citenamefont {Saintillan}\ and\ \citenamefont
  {Shelley}(2008)}]{Saintillan2008}%
  \BibitemOpen
  \bibfield  {author} {\bibinfo {author} {\bibfnamefont {D.}~\bibnamefont
  {Saintillan}}\ and\ \bibinfo {author} {\bibfnamefont {M.~J.}\ \bibnamefont
  {Shelley}},\ }\href {\doibase 10.1103/PhysRevLett.100.178103} {\bibfield
  {journal} {\bibinfo  {journal} {Physical Review Letters}\ }\textbf {\bibinfo
  {volume} {100}},\ \bibinfo {pages} {1} (\bibinfo {year} {2008})}\BibitemShut
  {NoStop}%
\bibitem [{\citenamefont {Doostmohammadi}\ \emph {et~al.}(2015)\citenamefont
  {Doostmohammadi}, \citenamefont {Adamer}, \citenamefont {Thampi},\ and\
  \citenamefont {Yeomans}}]{Doostmohammadi2015}%
  \BibitemOpen
  \bibfield  {author} {\bibinfo {author} {\bibfnamefont {A.}~\bibnamefont
  {Doostmohammadi}}, \bibinfo {author} {\bibfnamefont {M.}~\bibnamefont
  {Adamer}}, \bibinfo {author} {\bibfnamefont {S.~P.}\ \bibnamefont {Thampi}},
  \ and\ \bibinfo {author} {\bibfnamefont {J.~M.}\ \bibnamefont {Yeomans}},\
  }\href@noop {} {\bibfield  {journal} {\bibinfo  {journal} {arXiv.org}\ ,\
  \bibinfo {pages} {1}} (\bibinfo {year} {2015})},\ \Eprint
  {http://arxiv.org/abs/arXiv:1505.04199v1} {arXiv:arXiv:1505.04199v1}
  \BibitemShut {NoStop}%
\bibitem [{\citenamefont {Narayan}\ \emph {et~al.}(2007)\citenamefont
  {Narayan}, \citenamefont {Ramaswamy},\ and\ \citenamefont
  {Menon}}]{Narayan2007}%
  \BibitemOpen
  \bibfield  {author} {\bibinfo {author} {\bibfnamefont {V.}~\bibnamefont
  {Narayan}}, \bibinfo {author} {\bibfnamefont {S.}~\bibnamefont {Ramaswamy}},
  \ and\ \bibinfo {author} {\bibfnamefont {N.}~\bibnamefont {Menon}},\ }\href
  {\doibase 10.1126/science.1140414} {\bibfield  {journal} {\bibinfo  {journal}
  {Science}\ }\textbf {\bibinfo {volume} {317}},\ \bibinfo {pages} {105}
  (\bibinfo {year} {2007})}\BibitemShut {NoStop}%
\bibitem [{\citenamefont {Kemkemer}\ \emph {et~al.}(2000)\citenamefont
  {Kemkemer}, \citenamefont {Kling}, \citenamefont {Kaufmann},\ and\
  \citenamefont {Gruler}}]{Kemkemer2000}%
  \BibitemOpen
  \bibfield  {author} {\bibinfo {author} {\bibfnamefont {R.}~\bibnamefont
  {Kemkemer}}, \bibinfo {author} {\bibfnamefont {D.}~\bibnamefont {Kling}},
  \bibinfo {author} {\bibfnamefont {D.}~\bibnamefont {Kaufmann}}, \ and\
  \bibinfo {author} {\bibfnamefont {H.}~\bibnamefont {Gruler}},\ }\href@noop {}
  {\bibfield  {journal} {\bibinfo  {journal} {Eur. Phys. J. E}\ }\textbf
  {\bibinfo {volume} {1}},\ \bibinfo {pages} {215} (\bibinfo {year}
  {2000})}\BibitemShut {NoStop}%
\bibitem [{\citenamefont {Duclos}\ \emph {et~al.}(2014)\citenamefont {Duclos},
  \citenamefont {Garcia}, \citenamefont {Yevick},\ and\ \citenamefont
  {Silberzan}}]{Duclos2014}%
  \BibitemOpen
  \bibfield  {author} {\bibinfo {author} {\bibfnamefont {G.}~\bibnamefont
  {Duclos}}, \bibinfo {author} {\bibfnamefont {S.}~\bibnamefont {Garcia}},
  \bibinfo {author} {\bibfnamefont {H.~G.}\ \bibnamefont {Yevick}}, \ and\
  \bibinfo {author} {\bibfnamefont {P.}~\bibnamefont {Silberzan}},\ }\href
  {\doibase 10.1039/c3sm52323c} {\bibfield  {journal} {\bibinfo  {journal}
  {Soft matter}\ }\textbf {\bibinfo {volume} {10}},\ \bibinfo {pages} {2346}
  (\bibinfo {year} {2014})}\BibitemShut {NoStop}%
\bibitem [{\citenamefont {Simha}\ and\ \citenamefont
  {Ramaswamy}(2002)}]{Simha2002}%
  \BibitemOpen
  \bibfield  {author} {\bibinfo {author} {\bibfnamefont {R.}~\bibnamefont
  {Simha}}\ and\ \bibinfo {author} {\bibfnamefont {S.}~\bibnamefont
  {Ramaswamy}},\ }\href {\doibase 10.1016/S0378-4371(02)00503-4} {\bibfield
  {journal} {\bibinfo  {journal} {Physica A: Statistical Mechanics and its
  Applications}\ }\textbf {\bibinfo {volume} {306}},\ \bibinfo {pages} {262}
  (\bibinfo {year} {2002})}\BibitemShut {NoStop}%
\bibitem [{\citenamefont {Ramaswamy}\ \emph {et~al.}(2003)\citenamefont
  {Ramaswamy}, \citenamefont {Simha},\ and\ \citenamefont
  {Toner}}]{Ramaswamy2003}%
  \BibitemOpen
  \bibfield  {author} {\bibinfo {author} {\bibfnamefont {S.}~\bibnamefont
  {Ramaswamy}}, \bibinfo {author} {\bibfnamefont {R.~A.}\ \bibnamefont
  {Simha}}, \ and\ \bibinfo {author} {\bibfnamefont {J.}~\bibnamefont
  {Toner}},\ }\href {\doibase 10.1209/epl/i2003-00346-7} {\bibfield  {journal}
  {\bibinfo  {journal} {Europhys. Lett.}\ }\textbf {\bibinfo {volume} {62}},\
  \bibinfo {pages} {196} (\bibinfo {year} {2003})}\BibitemShut {NoStop}%
\bibitem [{\citenamefont {Toner}\ and\ \citenamefont {Tu}(1995)}]{Toner1995}%
  \BibitemOpen
  \bibfield  {author} {\bibinfo {author} {\bibfnamefont {J.}~\bibnamefont
  {Toner}}\ and\ \bibinfo {author} {\bibfnamefont {Y.}~\bibnamefont {Tu}},\
  }\href {\doibase 10.1103/PhysRevLett.75.4326} {\bibfield  {journal} {\bibinfo
   {journal} {Phys. Rev. Lett.}\ }\textbf {\bibinfo {volume} {75}},\ \bibinfo
  {pages} {4326} (\bibinfo {year} {1995})}\BibitemShut {NoStop}%
\bibitem [{\citenamefont {Toner}\ and\ \citenamefont {Tu}(1998)}]{Toner1998}%
  \BibitemOpen
  \bibfield  {author} {\bibinfo {author} {\bibfnamefont {J.}~\bibnamefont
  {Toner}}\ and\ \bibinfo {author} {\bibfnamefont {Y.}~\bibnamefont {Tu}},\
  }\href {\doibase 10.1103/PhysRevE.58.4828} {\bibfield  {journal} {\bibinfo
  {journal} {Phys. Rev. E}\ }\textbf {\bibinfo {volume} {58}},\ \bibinfo
  {pages} {4828} (\bibinfo {year} {1998})}\BibitemShut {NoStop}%
\bibitem [{\citenamefont {Pismen}(2013)}]{Pismen2013a}%
  \BibitemOpen
  \bibfield  {author} {\bibinfo {author} {\bibfnamefont {L.~M.}\ \bibnamefont
  {Pismen}},\ }\href {\doibase 10.1103/PhysRevE.88.050502} {\bibfield
  {journal} {\bibinfo  {journal} {Phys. Rev. E}\ }\textbf {\bibinfo {volume}
  {88}},\ \bibinfo {pages} {050502} (\bibinfo {year} {2013})}\BibitemShut
  {NoStop}%
\bibitem [{\citenamefont {Srivastava}\ and\ \citenamefont
  {Marchetti}(2015{\natexlab{a}})}]{Communication2015}%
  \BibitemOpen
  \bibfield  {author} {\bibinfo {author} {\bibfnamefont {P.}~\bibnamefont
  {Srivastava}}\ and\ \bibinfo {author} {\bibfnamefont {M.~C.}\ \bibnamefont
  {Marchetti}},\ }\href@noop {} {}\bibinfo {howpublished} {Private
  Commnication} (\bibinfo {year} {2015}{\natexlab{a}})\BibitemShut {NoStop}%
\bibitem [{\citenamefont {Srivastava}\ and\ \citenamefont
  {Marchetti}(2015{\natexlab{b}})}]{MarchMeeting2015}%
  \BibitemOpen
  \bibfield  {author} {\bibinfo {author} {\bibfnamefont {P.}~\bibnamefont
  {Srivastava}}\ and\ \bibinfo {author} {\bibfnamefont {M.~C.}\ \bibnamefont
  {Marchetti}},\ }\href@noop {} {\enquote {\bibinfo {title} {Instabilities and
  patterns in an active nematic film},}\ }\bibinfo {howpublished} {APS March
  Meeting} (\bibinfo {year} {2015}{\natexlab{b}})\BibitemShut {NoStop}%
\bibitem [{\citenamefont {Mishra}\ and\ \citenamefont
  {Ramaswamy}(2006)}]{Mishra2006}%
  \BibitemOpen
  \bibfield  {author} {\bibinfo {author} {\bibfnamefont {S.}~\bibnamefont
  {Mishra}}\ and\ \bibinfo {author} {\bibfnamefont {S.}~\bibnamefont
  {Ramaswamy}},\ }\href {\doibase 10.1103/PhysRevLett.97.090602} {\bibfield
  {journal} {\bibinfo  {journal} {Phys. Rev. Lett}\ }\textbf {\bibinfo {volume}
  {97}},\ \bibinfo {pages} {090602} (\bibinfo {year} {2006})}\BibitemShut
  {NoStop}%
\bibitem [{\citenamefont {Chat\'{e}}\ \emph {et~al.}(2006)\citenamefont
  {Chat\'{e}}, \citenamefont {Ginelli},\ and\ \citenamefont
  {Montagne}}]{Chate2006}%
  \BibitemOpen
  \bibfield  {author} {\bibinfo {author} {\bibfnamefont {H.}~\bibnamefont
  {Chat\'{e}}}, \bibinfo {author} {\bibfnamefont {F.}~\bibnamefont {Ginelli}},
  \ and\ \bibinfo {author} {\bibfnamefont {R.}~\bibnamefont {Montagne}},\
  }\href {\doibase 10.1103/PhysRevLett.96.180602} {\bibfield  {journal}
  {\bibinfo  {journal} {Phys. Rev. Lett}\ }\textbf {\bibinfo {volume} {96}},\
  \bibinfo {pages} {180602} (\bibinfo {year} {2006})}\BibitemShut {NoStop}%
\bibitem [{\citenamefont {Baskaran}\ and\ \citenamefont
  {Marchetti}(2008{\natexlab{a}})}]{Baskaran2008}%
  \BibitemOpen
  \bibfield  {author} {\bibinfo {author} {\bibfnamefont {A.}~\bibnamefont
  {Baskaran}}\ and\ \bibinfo {author} {\bibfnamefont {M.~C.}\ \bibnamefont
  {Marchetti}},\ }\href {\doibase 10.1103/PhysRevE.77.011920} {\bibfield
  {journal} {\bibinfo  {journal} {Physical Review E}\ }\textbf {\bibinfo
  {volume} {77}},\ \bibinfo {pages} {011920} (\bibinfo {year}
  {2008}{\natexlab{a}})}\BibitemShut {NoStop}%
\bibitem [{\citenamefont {Baskaran}\ and\ \citenamefont
  {Marchetti}(2008{\natexlab{b}})}]{Baskaran2008a}%
  \BibitemOpen
  \bibfield  {author} {\bibinfo {author} {\bibfnamefont {A.}~\bibnamefont
  {Baskaran}}\ and\ \bibinfo {author} {\bibfnamefont {M.}~\bibnamefont
  {Marchetti}},\ }\href {\doibase 10.1103/PhysRevLett.101.268101} {\bibfield
  {journal} {\bibinfo  {journal} {Phys. Rev. Lett}\ }\textbf {\bibinfo {volume}
  {101}},\ \bibinfo {pages} {268101} (\bibinfo {year}
  {2008}{\natexlab{b}})}\BibitemShut {NoStop}%
\bibitem [{\citenamefont {Mishra}\ \emph {et~al.}(2010)\citenamefont {Mishra},
  \citenamefont {{Aditi Simha}},\ and\ \citenamefont {Ramaswamy}}]{Mishra2010}%
  \BibitemOpen
  \bibfield  {author} {\bibinfo {author} {\bibfnamefont {S.}~\bibnamefont
  {Mishra}}, \bibinfo {author} {\bibfnamefont {R.}~\bibnamefont {{Aditi
  Simha}}}, \ and\ \bibinfo {author} {\bibfnamefont {S.}~\bibnamefont
  {Ramaswamy}},\ }\href {\doibase 10.1088/1742-5468/2010/02/P02003} {\bibfield
  {journal} {\bibinfo  {journal} {J. Stat. Mech. Theory Exp.}\ }\textbf
  {\bibinfo {volume} {2010}},\ \bibinfo {pages} {P02003} (\bibinfo {year}
  {2010})}\BibitemShut {NoStop}%
\bibitem [{\citenamefont {Peruani}\ \emph {et~al.}(2011)\citenamefont
  {Peruani}, \citenamefont {Ginelli}, \citenamefont {B\"{a}r},\ and\
  \citenamefont {Chat\'{e}}}]{Peruani2011}%
  \BibitemOpen
  \bibfield  {author} {\bibinfo {author} {\bibfnamefont {F.}~\bibnamefont
  {Peruani}}, \bibinfo {author} {\bibfnamefont {F.}~\bibnamefont {Ginelli}},
  \bibinfo {author} {\bibfnamefont {M.}~\bibnamefont {B\"{a}r}}, \ and\
  \bibinfo {author} {\bibfnamefont {H.}~\bibnamefont {Chat\'{e}}},\ }\href
  {\doibase 10.1088/1742-6596/297/1/012014} {\bibfield  {journal} {\bibinfo
  {journal} {J. Phys. Conf. Ser.}\ }\textbf {\bibinfo {volume} {297}},\
  \bibinfo {pages} {012014} (\bibinfo {year} {2011})}\BibitemShut {NoStop}%
\bibitem [{\citenamefont {Peshkov}\ \emph
  {et~al.}(2012{\natexlab{a}})\citenamefont {Peshkov}, \citenamefont {Ngo},
  \citenamefont {Bertin}, \citenamefont {Chat\'{e}},\ and\ \citenamefont
  {Ginelli}}]{Peshkov2012}%
  \BibitemOpen
  \bibfield  {author} {\bibinfo {author} {\bibfnamefont {A.}~\bibnamefont
  {Peshkov}}, \bibinfo {author} {\bibfnamefont {S.}~\bibnamefont {Ngo}},
  \bibinfo {author} {\bibfnamefont {E.}~\bibnamefont {Bertin}}, \bibinfo
  {author} {\bibfnamefont {H.}~\bibnamefont {Chat\'{e}}}, \ and\ \bibinfo
  {author} {\bibfnamefont {F.}~\bibnamefont {Ginelli}},\ }\href {\doibase
  10.1103/PhysRevLett.109.098101} {\bibfield  {journal} {\bibinfo  {journal}
  {Phys. Rev. Lett.}\ }\textbf {\bibinfo {volume} {109}},\ \bibinfo {pages}
  {098101} (\bibinfo {year} {2012}{\natexlab{a}})}\BibitemShut {NoStop}%
\bibitem [{\citenamefont {Peshkov}\ \emph
  {et~al.}(2012{\natexlab{b}})\citenamefont {Peshkov}, \citenamefont {Aranson},
  \citenamefont {Bertin}, \citenamefont {Chat\'{e}},\ and\ \citenamefont
  {Ginelli}}]{Peshkov2012a}%
  \BibitemOpen
  \bibfield  {author} {\bibinfo {author} {\bibfnamefont {A.}~\bibnamefont
  {Peshkov}}, \bibinfo {author} {\bibfnamefont {I.~S.}\ \bibnamefont
  {Aranson}}, \bibinfo {author} {\bibfnamefont {E.}~\bibnamefont {Bertin}},
  \bibinfo {author} {\bibfnamefont {H.}~\bibnamefont {Chat\'{e}}}, \ and\
  \bibinfo {author} {\bibfnamefont {F.}~\bibnamefont {Ginelli}},\ }\href
  {\doibase 10.1103/PhysRevLett.109.268701} {\bibfield  {journal} {\bibinfo
  {journal} {Phys. Rev. Lett.}\ }\textbf {\bibinfo {volume} {109}},\ \bibinfo
  {pages} {268701} (\bibinfo {year} {2012}{\natexlab{b}})}\BibitemShut
  {NoStop}%
\bibitem [{\citenamefont {Baskaran}\ and\ \citenamefont
  {Marchetti}(2012)}]{Baskaran2012}%
  \BibitemOpen
  \bibfield  {author} {\bibinfo {author} {\bibfnamefont {A.}~\bibnamefont
  {Baskaran}}\ and\ \bibinfo {author} {\bibfnamefont {M.~C.}\ \bibnamefont
  {Marchetti}},\ }\href {\doibase 10.1140/epje/i2012-12095-8} {\bibfield
  {journal} {\bibinfo  {journal} {Eur. Phys. J. E.}\ }\textbf {\bibinfo
  {volume} {35}},\ \bibinfo {pages} {95} (\bibinfo {year} {2012})}\BibitemShut
  {NoStop}%
\bibitem [{\citenamefont {Bertin}\ \emph {et~al.}(2013)\citenamefont {Bertin},
  \citenamefont {Chat\'{e}}, \citenamefont {Ginelli}, \citenamefont {Mishra},
  \citenamefont {Peshkov},\ and\ \citenamefont {Ramaswamy}}]{Bertin2013}%
  \BibitemOpen
  \bibfield  {author} {\bibinfo {author} {\bibfnamefont {E.}~\bibnamefont
  {Bertin}}, \bibinfo {author} {\bibfnamefont {H.}~\bibnamefont {Chat\'{e}}},
  \bibinfo {author} {\bibfnamefont {F.}~\bibnamefont {Ginelli}}, \bibinfo
  {author} {\bibfnamefont {S.}~\bibnamefont {Mishra}}, \bibinfo {author}
  {\bibfnamefont {A.}~\bibnamefont {Peshkov}}, \ and\ \bibinfo {author}
  {\bibfnamefont {S.}~\bibnamefont {Ramaswamy}},\ }\href {\doibase
  10.1088/1367-2630/15/8/085032} {\bibfield  {journal} {\bibinfo  {journal}
  {New J. Phys.}\ }\textbf {\bibinfo {volume} {15}},\ \bibinfo {pages} {085032}
  (\bibinfo {year} {2013})}\BibitemShut {NoStop}%
\bibitem [{\citenamefont {Ngo}\ \emph {et~al.}(2013)\citenamefont {Ngo},
  \citenamefont {Peshkov}, \citenamefont {Aranson}, \citenamefont {Bertin},
  \citenamefont {Ginelli},\ and\ \citenamefont {Chat\'{e}}}]{Ngo2013}%
  \BibitemOpen
  \bibfield  {author} {\bibinfo {author} {\bibfnamefont {S.}~\bibnamefont
  {Ngo}}, \bibinfo {author} {\bibfnamefont {A.}~\bibnamefont {Peshkov}},
  \bibinfo {author} {\bibfnamefont {I.~S.}\ \bibnamefont {Aranson}}, \bibinfo
  {author} {\bibfnamefont {E.}~\bibnamefont {Bertin}}, \bibinfo {author}
  {\bibfnamefont {F.}~\bibnamefont {Ginelli}}, \ and\ \bibinfo {author}
  {\bibfnamefont {H.}~\bibnamefont {Chat\'{e}}},\ }\href {\doibase
  10.1103/PhysRevLett.113.038302} {\bibfield  {journal} {\bibinfo  {journal}
  {Physical Review Letters}\ ,\ \bibinfo {pages} {038302}} (\bibinfo {year}
  {2013})},\ \Eprint {http://arxiv.org/abs/1312.1076} {1312.1076} \BibitemShut
  {NoStop}%
\bibitem [{\citenamefont {Putzig}\ and\ \citenamefont
  {Baskaran}(2014)}]{Putzig2014}%
  \BibitemOpen
  \bibfield  {author} {\bibinfo {author} {\bibfnamefont {E.}~\bibnamefont
  {Putzig}}\ and\ \bibinfo {author} {\bibfnamefont {A.}~\bibnamefont
  {Baskaran}},\ }\href {\doibase 10.1103/PhysRevE.90.042304} {\bibfield
  {journal} {\bibinfo  {journal} {Physical Review E}\ }\textbf {\bibinfo
  {volume} {90}},\ \bibinfo {pages} {042304} (\bibinfo {year} {2014})},\
  \Eprint {http://arxiv.org/abs/1057984} {1057984} \BibitemShut {NoStop}%
\bibitem [{Note1()}]{Note1}%
  \BibitemOpen
  \bibinfo {note} {Details of the this transition, as well as the additional
  discussion concerning the equations and linear stability analysis, can be
  found in the supplement.}\BibitemShut {Stop}%
\bibitem [{\citenamefont {Giomi}\ \emph {et~al.}(2011)\citenamefont {Giomi},
  \citenamefont {Mahadevan}, \citenamefont {Chakraborty},\ and\ \citenamefont
  {Hagan}}]{Giomi2011}%
  \BibitemOpen
  \bibfield  {author} {\bibinfo {author} {\bibfnamefont {L.}~\bibnamefont
  {Giomi}}, \bibinfo {author} {\bibfnamefont {L.}~\bibnamefont {Mahadevan}},
  \bibinfo {author} {\bibfnamefont {B.}~\bibnamefont {Chakraborty}}, \ and\
  \bibinfo {author} {\bibfnamefont {M.~F.}\ \bibnamefont {Hagan}},\ }\href
  {\doibase 10.1103/PhysRevLett.106.218101} {\bibfield  {journal} {\bibinfo
  {journal} {Phys. Rev. Lett.}\ }\textbf {\bibinfo {volume} {106}},\ \bibinfo
  {pages} {218101} (\bibinfo {year} {2011})}\BibitemShut {NoStop}%
\bibitem [{\citenamefont {Giomi}\ \emph {et~al.}(2012)\citenamefont {Giomi},
  \citenamefont {Mahadevan}, \citenamefont {Chakraborty},\ and\ \citenamefont
  {Hagan}}]{Giomi2012}%
  \BibitemOpen
  \bibfield  {author} {\bibinfo {author} {\bibfnamefont {L.}~\bibnamefont
  {Giomi}}, \bibinfo {author} {\bibfnamefont {L.}~\bibnamefont {Mahadevan}},
  \bibinfo {author} {\bibfnamefont {B.}~\bibnamefont {Chakraborty}}, \ and\
  \bibinfo {author} {\bibfnamefont {M.~F.}\ \bibnamefont {Hagan}},\ }\href@noop
  {} {\bibfield  {journal} {\bibinfo  {journal} {Nonlinearity}\ ,\ \bibinfo
  {pages} {2245}} (\bibinfo {year} {2012})}\BibitemShut {NoStop}%
\end{thebibliography}%


%
\end{document}


\title{ Supplementary Information \\
\small (Instabilities, defects, and defect ordering in an overdamped active nematic) }

\author{Elias Putzig}
\email[]{efputzig@brandeis.edu}
\affiliation{Martin Fisher School of Physics, Brandeis University, Waltham, MA 02453, USA}
\author{Gabriel S. Redner}
\email[]{gredner@brandeis.edu}
\affiliation{Martin Fisher School of Physics, Brandeis University, Waltham, MA 02453, USA}
\author{Arvind Baskaran}
\email[]{baskaran@math.uci.edu}
\affiliation{Department of Chemistry and Biochemistry, University of Maryland, College Park, MD 20742, USA}
\author{Aparna Baskaran}
\email[]{aparna@brandeis.edu}
\affiliation{Martin Fisher School of Physics, Brandeis University, Waltham, MA 02453, USA}

\maketitle

\section{ Free Energy } \label{LdGF}

The dynamics of an equilibrium liquid crystal, which were discussed in the body of the text, can be derived from gradient descent on the well-known Landau-de Gennes free energy \cite{deGennes1993}:
\beq \label{FreeEnergy}
\begin{split}
F = \int d^{d}r \Big[ &- \frac{\alpha}{2}Tr({\bf Q}^{2}) 
+\frac{\beta}{4}\big(Tr({\bf Q}^{2})\big)^{2}  
+ \chi Q_{ij}\partial_{i}\partial_{j}\rho
+ \frac{1}{2}\kappa Q_{ij}\nabla^4Q_{ij}\\
& + \frac{1}{2}L_1(\partial_{i}Q_{kl})^{2} 
+\frac{1}{2}L_2(\partial_{i}Q_{ik})\partial_{j}Q_{jk} 
+ \frac{1}{2\rho}L_{4}Q_{ij}(\partial_{i}Q_{kl})\partial_{j}Q_{kl} \Big]  +\Phi[\rho] 
\end{split}
\eeq
where the homogeneous terms give a second order phase transition and the terms with coefficients $L_{i}$ are elastic terms, associated with the energy cost of director distortions. The elastic energy cost is widely described in terms of splay ($K_{11}(\nabla\cdot \hat{n})^2$) and bend ($K_{33}\big(\hat{n}\times(\nabla\times\hat{n})\big)^{2}$) distortion of the director. The parameters in (\ref{FreeEnergy}) are related to the splay and bend coefficients as: $K_{11}=S^2(2L_1+L_2-SL_4)$ and $K_{33}=S^2(2L_1+L_2+2SL_4)$. 

The gradient-descent dynamics from the body of the paper came from taking a derivative of Eq. \ref{FreeEnergy}:
\beqs
  \label{Qfull}
  \begin{split}
-\gamma^{-1} \frac{\delta F}{\delta Q_{ij}} 
  & = D_r[\alpha-\beta Tr{\bf Q}^{2}] Q_{ij} 
  + 2D_E \nabla^{2} Q_{ij}  
  +D_{\rho}(\partial_{i}\partial_{j} - \frac{1}{2}\delta_{ij}\nabla^{2})\rho 
  - K\nabla^4Q_{ij}\\
  & + \frac{D_{\delta}}{\rho} \Big( 2Q_{kl}\partial_{k}\partial_{l}Q_{ij} 
  +2 [\partial_{k}Q_{kl}]\partial_{l}Q_{ij}
  -\big([\partial_{i}Q_{kl}]\partial_{j}Q_{kl} 
  -\frac{1}{2}\delta_{ij}[\partial_{k}Q_{lm}]^{2}\big)   \Big )
  \end{split}
\eeqs
where $D_{r}=2/\gamma$, $D_{\rho}=-2 \chi /\gamma$, $K=2 \kappa /\gamma$ and the two mean elastic terms have been consolidated by using the tracelessness and symmetry of $Q$ to group them in the term with coefficient $D_E=2L_{1}/\gamma=2L_{2}/\gamma$. The differential elastic coefficient, $D_{\delta}=2L_{4}/\gamma$, has a range $-\frac{3}{2S}<\frac{D_\delta}{D_E}<\frac{3}{S}$, and is set by the requirement that the equilibrium nematic should be stable to director fluctuations. Further a fourth order gradient term (with coefficient $K$) is included in order to ensure smoothness and numerical stability.

\section{ Active stresses and self-induced flows} \label{Flows}

In the body of the text we considered the dynamics of the director when the system was in an imposed flow,in which case the dynamical equation becomes
\begin{gather*}
(\partial_{t}+\vec{u}\cdot\vec{\nabla})Q_{ij}=-\gamma^{-1}\frac{\delta F}{\delta Q_{ij}}+(Q_{ik}\Omega_{kj}-\Omega_{ik}Q_{kj})\\
+\lambda E_{ij}^{\mathcal{T}}+\lambda(Q_{ik}E_{kj}+E_{ik}Q_{kj})^{\mathcal{T}}
\end{gather*}
where $\Omega_{ij}$ and $E_{ij}$ are the vorticity and strain-rate tensor associated with the flow, and $\lambda$ is the flow-alignment parameter. We then postulated a form for the self-generated flow
 $\rho\vec{u}=-\lambda_{C}\vec{\nabla {\bf Q}}$, and the vorticity
$\rho\vec{\omega}=\lambda_{R}(\vec{\nabla}\times\vec{\nabla {\bf Q}})$.  This is consistent with a stress proportional to the nematic order, $\boldsymbol{\sigma}=\sigma_a{\bf Q}$ where coefficients ($\sigma_a$, $\lambda_C$ and $\lambda_R$) are positive for systems of extensile, `pusher' particles, and negative for contractile, `puller' particles.

We construct the tensors $\Omega_{ij}$ and $E_{ij}$ in a straightforward manner from these flows. The z- component of the vorticity vector gives the amount that
a fluid element is rotated by the self-generated flow in two dimensions, and the vorticity tensor,
which comes into the dynamics, is $\rho\Omega_{ij}=\frac{1}{2}\epsilon_{ijk}\omega_{k}$
where $\epsilon$ is the Levi-Civita symbol. The strain-rate tensor, which is the symmetrized gradient of the flow 
$E_{ij}=\frac{1}{2}(\partial_iu_j-\partial_ju_i)$
will play a role if the particles tend to align or tumble in shear.

We postulated that an active liquid crystal undergoes gradient descent
dynamics so as to minimize the energy cost of director distortions
and inhomogeneities, in the local rest frame of this self-generated
flow arising from the activity. Using the form of the flow and vorticity
arising from the activity identified above, our dynamical equations
for an active nematic take the form
\begin{gather} \label{QDynam}
\partial_{t}Q_{ij} =-\gamma^{-1}\frac{\delta F}{\delta Q_{ij}} 
+\frac{\lambda_{C}}{\rho}[\partial_{k}Q_{kl}]\partial_{l}Q_{ij}
-\frac{\lambda_{S}}{2\rho}Q_{ij}\partial_{k}\partial_{l}Q_{kl} \\
- \frac{\lambda_{R}}{2\rho}\big(Q_{i k}\partial_{k}\partial_{l}Q_{j l}-Q_{i k}\partial_{j}\partial_{l}Q_{kl}-Q_{kj}\partial_{i}\partial_{l}Q_{kl}+Q_{kj}\partial_{k}\partial_{l}Q_{i l}\big)\nonumber
\end{gather}
where $\lambda_{S}=\lambda\lambda_{C}$ is the strength of
flow-alignment, and we have simplified the flow-alignment term in the expression 
above using the symmetry and tracelessness of the nematic order tensor.

The form of the flow-alignment term in this overdamped case is different from the form that arises in the overdamped limit of the the theories in which flow arises through a Stokes equation (such as \cite{Giomi2013, Thampi2013, Blow2014, Giomi2014, Thampi2014a, Thampi2014, Thampi2014b}).  This is because the phenomenological, overdamped theory which we consider here does not have incompressible flows.  Because of this, the first order flow-alignment term in the gradient descent dynamics ($\lambda E_{ij}^{\mathcal{T}}$) is not intrinsically traceless and must be made so by hand in order to preserve the symmetric and traceless nature of ${\bf Q}$, and the second order term ($\lambda(Q_{ik}E_{kj}+E_{ik}Q_{kj})^{\mathcal{T}}$) is nonzero, whereas it is zero in two dimensions for incompressible flow.  The traceless version of the first order flow-alignment term is included in the functional derivative in Eq. \ref{QDynam}, as it is the same as the mean elastic term: $E_{ij}^{\mathcal{T}}=\partial_{k}(\partial_{i}Q_{kj} +\partial_{j}Q_{ik}-\delta_{ij}\partial_{l}Q_{kl})=\nabla^2Q_{ij}$.

\section{ Linear Stability } \label{LSA}

In the homogeneous limit, Eq. (\ref{QDynam}) and the dynamical equation for the density,
$\partial_t\rho=D\nabla^2\rho+D_Q\boldsymbol{\nabla \nabla}:{\bf Q}$, 
admit a homogeneous uni-axial nematic state 
with average density $\rho_0>1$, and the order parameter ${\bf Q}=\rho_0S_0(\boldsymbol{\hat{x}\hat{x}}-\frac{1}{2}{\bf I})$ with  $S_0=\sqrt{\frac{2(\rho_0-1)}{\rho_0+1}}$.  Without loss of generality, we have picked coordinates so that the direction of nematic ordering is along the $x$-axis of our coordinate system. We parameterize the fluctuations about this state in the form
 $\rho(\mathbf{r},t)=\rho_0+\delta\rho(\mathbf{r},t)$, $Q_{xx}(\mathbf{r},t)=\frac{1}{2}\rho_0S_{0}+\delta Q_{xx}(\mathbf{r},t)$ and $Q_{xy}(\mathbf{r},t)=\delta Q_{xy}(\mathbf{r},t)$. Further, we introduce a Fourier transform
$\tilde{X} = \int d\mathbf{r}  e^{ i\mathbf{k}\cdot \mathbf{r} }X(\mathbf{r},t)$. The resulting linearized equations in Fourier space are
\begin{subequations}
\beq \label{deltarho}
  \partial_{t}\delta\tilde{\rho} =
- Dk^{2}\delta\tilde{\rho} -D_{Q}k^{2}\cos(2\phi)\delta\tilde{Q}_{xx} 
- D_{Q}k^{2}\sin(2\phi)\delta\tilde{Q}_{xy}
\eeq
\beq \label{deltaQxx}
\begin{array}{rcl}
  \partial_{t}\delta\tilde{Q}_{xx}
  &=& \Big(C_{0}-\frac{1}{2}D_{\rho}k^{2}\cos(2\phi)\Big) \delta\tilde{\rho} \\
  &-& \Big(2\alpha_{0}+2D_{E}k^{2} +(D_{\delta}-\frac{1}{2}\lambda_S) S_{0}k^{2}\cos(2\phi)
  +Kk^4 \Big)\delta\tilde{Q}_{xx} \\
  &+&\frac{1}{2}\lambda_SS_0k^2\sin(2\phi)\delta\tilde{Q}_{xy}
\end{array}
\eeq
\beq \label{deltaQxy}
\begin{array}{rcl}
  \partial_{t}\delta\tilde{Q}_{xy} =
  &-& \frac{1}{2}D_{\rho}k^{2}\sin(2\phi)\delta\tilde{\rho}  \\ &-&\frac{\lambda_{R}}{2}S_{0}k^{2}\sin(2\phi)\delta\tilde{Q}_{xx} \\
  &- &\Big(2D_{E}k^2-\frac{\lambda_{R}}{2}S_{0}k^2\cos(2\phi) 
  +D_{\delta}S_{0}k^2\cos(2\phi) +Kk^4 \Big) \delta\tilde{Q}_{yx} 
\end{array}
\eeq
\end{subequations}
where $\phi$ is the angle between the director (along the $x$ axis in our coordinates) and the spatial gradient vector
$\mathbf{k}$. Also, we have defined parameters $C_{0}=\sqrt{\frac{2(\rho_{0}-1)}{\rho_{0}+1}}\Big(\frac{\rho_{0}^{2}+\rho_{0}-1}{\rho_{0}+1}\Big)$,
and $\alpha_{0}=\rho_{0}-1$ for notational compactness. 

The stability of these equations is best explicated by considering spatial fluctuations along ($\phi=0$) or perpendicular to ($\phi=\frac{\pi}{2}$) the direction of ordering. In these two sectors, the fluctuations in the direction of ordering $\delta \tilde{Q}_{xy}$ decouple form those of the magnitude of ordering $\delta \tilde{Q}_{xx}$ and the density $\delta \tilde{\rho}$, thereby enabling a clear identification of mechanisms at play.

The decoupled dynamics of $\delta\tilde{Q}_{xy}$ are of the following form:
\beq \label{bendinstab}
\partial_{t}\delta\tilde{Q}_{xy} = -\Big( 2D_E(1\mp \psi)k^2 +Kk^4\Big)\delta\tilde{Q}_{xy}
\eeq
where the coefficient of $k^2$ is proportional to $(1-\psi)$ for fluctuations along the director ($\phi=\frac{\pi}{2}$) and $(1+\psi)$ for fluctuations which are perpendicular ($\phi=90^\circ$). The stability parameter, which is useful for describing the dynamics when bend in unstable, is $\psi = \frac{\lambda_R-2D_\delta}{D_E\cdot \Lambda (\rho_0)}$, where $\Lambda (\rho_0)=4\sqrt{\frac{\rho_0+1}{2(\rho_0-1)}}$. This instability is the primary mechanism for defect generation and formation of inhomogeneous steady states in this active system, as was discussed in the body of the paper.

Note that if the nematic was contractile, $\lambda_R$ is negative and hence there exists a splay instability, which occurs when $\psi < -1 $.  The analogous instability in fluid-mediated dynamics, which arises due to contractile stresses in the Stokes equation, has been studied in some detail  \cite{Giomi2014, Thampi2014, Thampi2014b}.  This splay instability leads to defect formation, despite the fact that contractile systems enhance defect annihilation \cite{Giomi2013}. We have, however, focused on extensile systems for this study of the overdamped dynamics, and left the study of the contractile, overdamped dynamics to future work.

\section{ Additional Numerical Phenomenology } \label{NumPhen}

\begin{figure}[hb]
  \includegraphics[width= \textwidth]{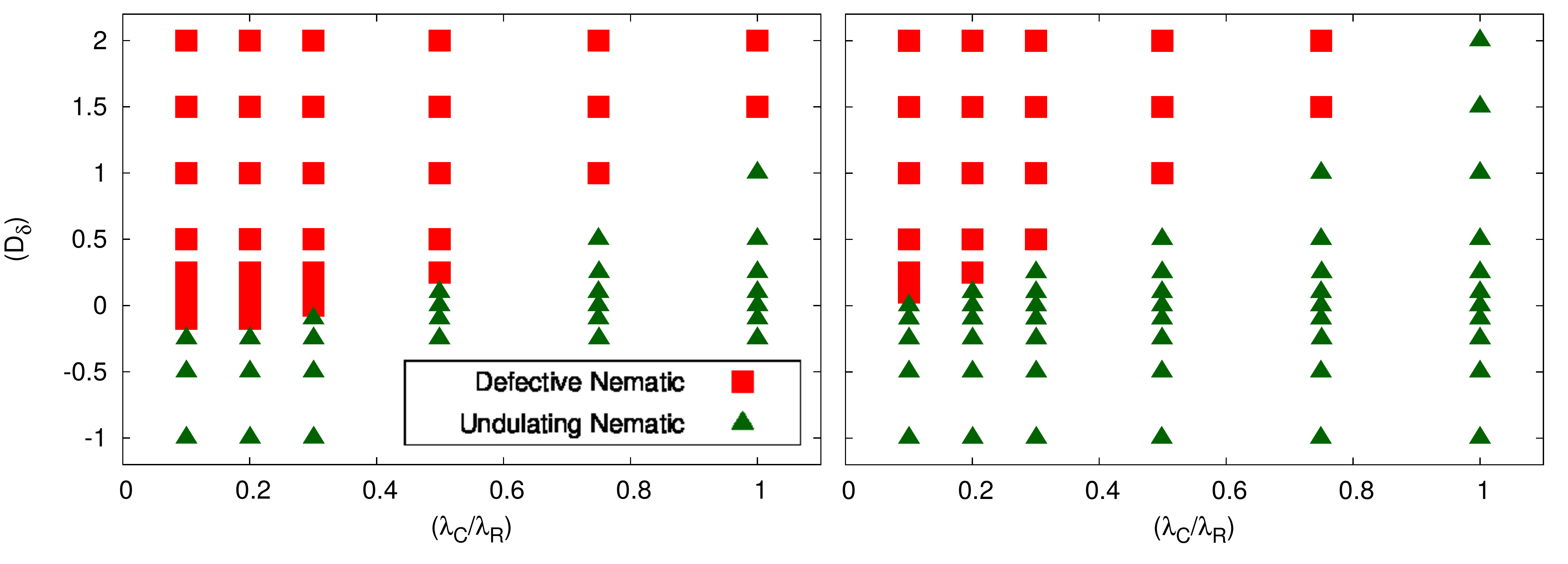}
  \caption{ These plots indicate whether an undulating nematic, or a turbulent, defective nematic state formed out of nematic initial conditions, from a point near the bend instability ($\psi=1.05$, $\rho_0=2.0$). Left Panel: A flow-tumbling system, with $\lambda_S=-1.0$. Right Panel: A flow-aligning system with $\lambda_S=1.0$.  The formation of defects near the critical density is facilitated by having a strength of active convection which is small compared to the strength of the active torque ($\lambda_C/\lambda_R<1$), and by a large energetic penalty to bend deformations, compared to the penalty to splay deformations ($D_\delta>0$).  Flow-tumbling systems also facilitate the formation of defects near the critical density, but flow-aligning versus flow-tumbling seems to be a small effect.
}
\label{UndVSDef}
\end{figure}

\begin{figure}[h]
  \includegraphics[width= 0.6\textwidth]{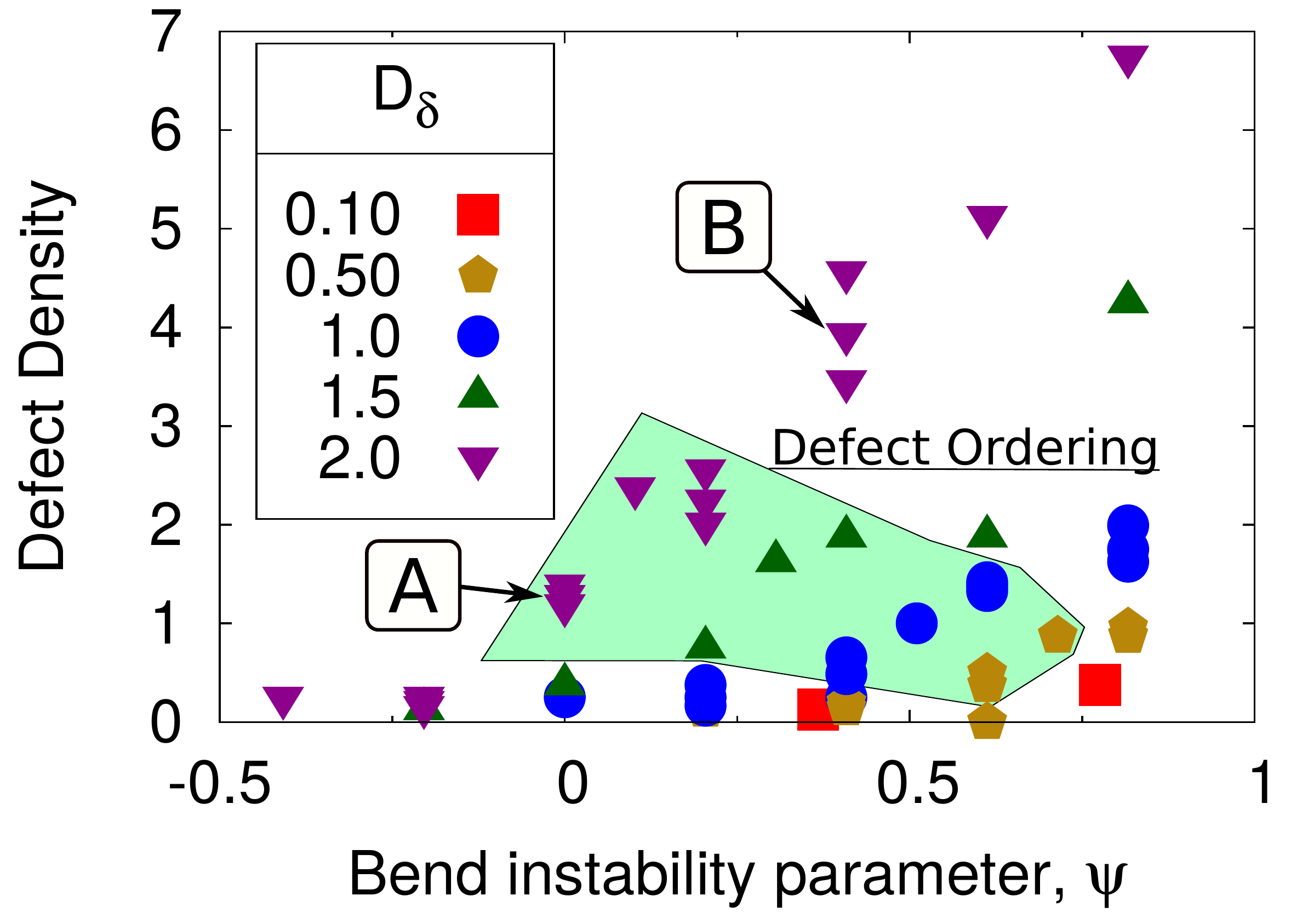}
  \caption{ This plot shows defect density (defects per $(100\ell_D)^2$), as a function of the bend instability parameter, below the bend instability ($\psi<1$). The shaded region shows where $+\frac{1}{2}$ defects showed statistically significant polar ordering.  This plot is a zoomed-in version of the plot on the left in Fig. 2 of the paper, and has parameters $\lambda_C=1.0$, and $\rho_0=2.0$. The two points indicated are systems which are shown in Fig. 1 in the body of the paper.  A is the defect-ordered state in the upper left, and B is the defective state below the bend instability, in the lower left.
}
\label{PolarOrder}
\end{figure}

\begin{figure}[h]
  \includegraphics[width= 0.6\textwidth]{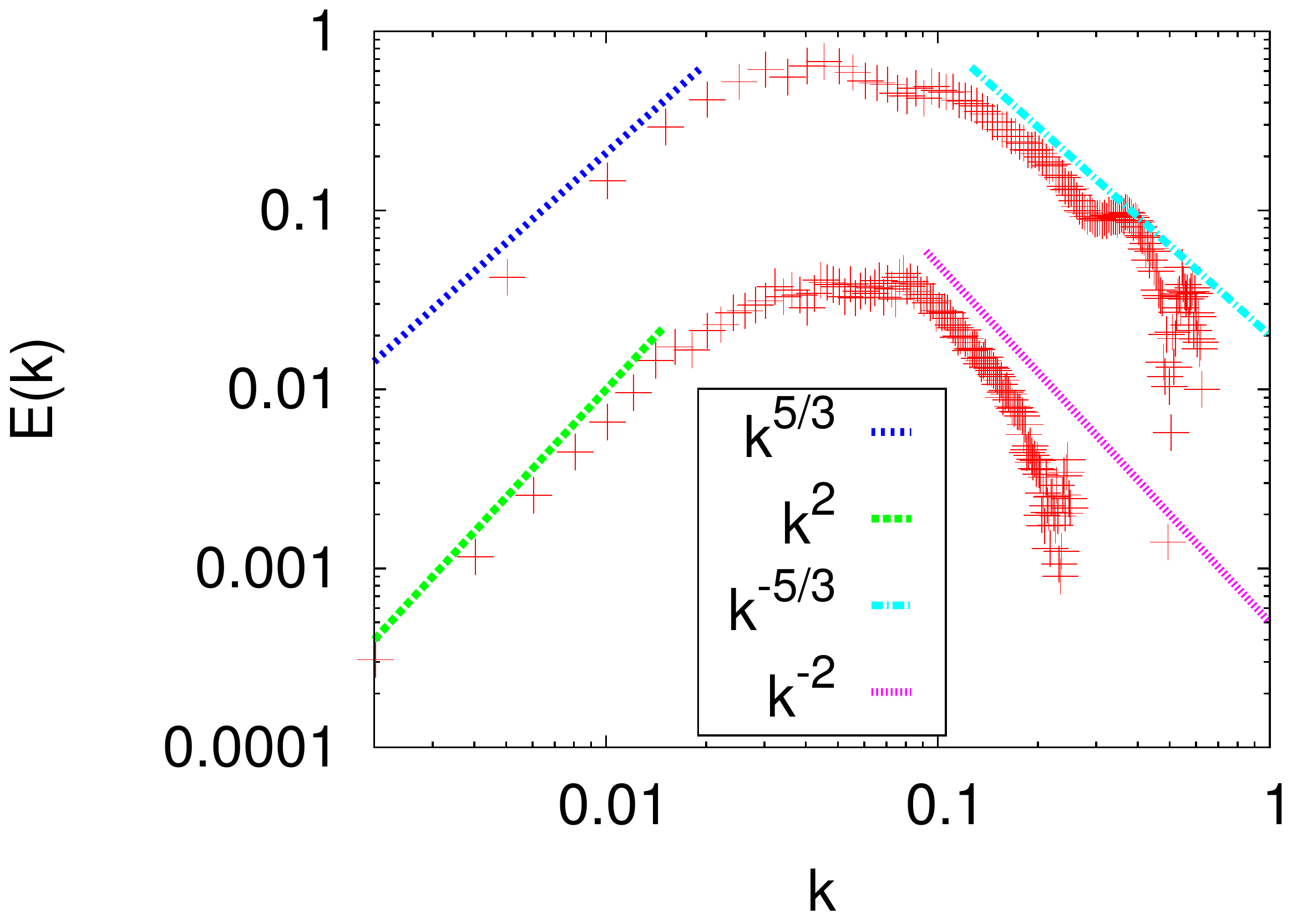}
  \caption{ Log-scale plots of the following approximation of the energy spectral density 
$E(k) = \frac{k}{2\pi}\int d^2Re^{ik\cdot R}\langle \vec{v}(0) \cdot \vec{v}(R) \rangle$.
Data shows two sets of parameters.
Top: $(\rho_0,\lambda_R,D_\delta)$=$(2,8,1)$ and 
Bottom: $(3,10,0.5)$; scaled by multiplicative constants for comparison. 
The energy density seems to scale with a power laws of $k^{5/3}$ ($k^{2}$) for small wavevector, and $k^{-5/3}$ ($k^{-2}$) and  for large wavevector for the upper (lower) curve. These u-shaped energy spectra are similar to what is seen in mixing flow in three dimensions, and is reminiscent of that seen in bacterial turbulence and a reversing-rod model \cite{Wensink2012}. 
}
\label{EnergySpectra}
\end{figure}

\bibliography{References}